\documentstyle[sprocltasi,equations]{article}

\input psfig

\def\smallfrac#1#2{\hbox{${\scriptstyle#1} \over {\scriptstyle#2}$}}
\def\fourth{{\scriptstyle{1 \over 4}}}
\def\half{{\scriptstyle{1\over 2}}}
\def\st{\scriptstyle}

\def\be{\begin{equation}}
\def\ee{\end{equation}}
\def\bea{\begin{eqnarray}}
\def\eea{\end{eqnarray}}
\def\eeq{\end{equation}}
\def\beq{\begin{equation}}

\def\@citex[#1]#2{\if@filesw\immediate\write\@auxout{\string\citation{#2}}\fi
  \@tempcnta\z@\@tempcntb\m@ne\def\@citea{}\@cite{\@for\@citeb:=#2\do
    {\@ifundefined
       {b@\@citeb}{\@citeo\@tempcntb\m@ne\@citea\def\@citea{,}{\bf ?}\@warning
       {Citation `\@citeb' on page \thepage \space undefined}}%
    {\setbox\z@\hbox{\global\@tempcntc0\csname b@\@citeb\endcsname\relax}%
     \ifnum\@tempcntc=\z@ \@citeo\@tempcntb\m@ne
       \@citea\def\@citea{,}\hbox{\csname b@\@citeb\endcsname}%
     \else
      \advance\@tempcntb\@ne
      \ifnum\@tempcntb=\@tempcntc
      \else\advance\@tempcntb\m@ne\@citeo
      \@tempcnta\@tempcntc\@tempcntb\@tempcntc\fi\fi}}\@citeo}{#1}}
\def\@citeo{\ifnum\@tempcnta>\@tempcntb\else\@citea\def\@citea{,}%
  \ifnum\@tempcnta=\@tempcntb\the\@tempcnta\else
   {\advance\@tempcnta\@ne\ifnum\@tempcnta=\@tempcntb \else \def\@citea{--}\fi
    \advance\@tempcnta\m@ne\the\@tempcnta\@citea\the\@tempcntb}\fi\fi}

\begin{document}
%
\def\dslash{\not{\hbox{\kern-2pt $\partial$}}}
\def\Dslash{\not{\hbox{\kern-4pt $D$}}}
\def\Oslash{\not{\hbox{\kern-4pt $O$}}}
\def\Qslash{\not{\hbox{\kern-4pt $Q$}}}
\def\pslash{\not{\hbox{\kern-2.3pt $p$}}}
\def\kslash{\not{\hbox{\kern-2.3pt $k$}}}
\def\qslash{\not{\hbox{\kern-2.3pt $q$}}}
 \newtoks\slashfraction
 \slashfraction={.13}
 \def\slash#1{\setbox0\hbox{$ #1 $}
 \setbox0\hbox to \the\slashfraction\wd0{\hss \box0}/\box0 }
 
 
%

\title{Supersymmetry Phenomenology (With a Broad Brush)}

\author{ Michael Dine }

\address{Santa Cruz Institute for Particle Physics \\
University of California, Santa Cruz, CA   95064}


\maketitle\abstracts{
These lectures provide an introduction to supersymmetry phenomenology.
They include an overview of the Minimal Supersymmetric Standard
Model.  The notion of soft breaking is explained, constraints
on the standard soft breaking parameters are reviewed,
and the standard ansatz of universal soft masses is discussed.
The rest of the lectures are devoted to understanding
supersymmetry breaking more microscopically.  Models of
dynamical supersymmetry breaking are reviewed, after which
we turn to the question of the scale of supersymmetry breaking.
Both intermediate and low scales and their phenomenology
are considered.  Finally, we consider applications to string
theory.  The emphasis, throughout, is on general issues
rather then extensive detail, but it is hoped that the listeners/readers
are prepared after these lectures to delve into the details.
}

\def\mhu{m_{H_U}^2}

\newcommand{\Meleven}{M_{11}}
\newcommand{\lp}{l_{11}}
\newcommand{\Releven}{R_{11}}

\section{Introduction}

The standard model is extraordinarily successful.  While
theorists (and sometimes
even experimentalists)
have become excited from time to time
about small discrepancies between data and theory,
at the moment of this writing, it is probably fair to say
that wherever comparison of theory and experiment is
possible, the agreement ranges between very good
to extremely good.
We are in the position, as we approach the twenty first
century, that we know with great
confidence all of the laws of nature which
operate down to distances at least as small as $10^{-16}$
cm.  

Yet we are firmly convinced that this structure is incomplete.
The theory has too many parameters.  It contains fundamental
scalars, something difficult to reconcile with our current
understanding of field theory.  Finally, it does not incorporate
gravity.  It is tempting to speculate that a new, as yet
undiscovered symmetry, supersymmetry, has something
to do with the answers to these questions.  Supersymmetry
is the only framework in which we seem to be able
to understand light fundamental scalars, i.e., scalars which
are pointlike down to scales much smaller than their
Compton wavelength.  Supersymmetry addresses the question
of parameters:   first, unification of gauge couplings
works much better with than without supersymmetry;
second,
it is easier to attack questions such as
fermion masses in supersymmetric theories, in
part simply due to the presence of fundamental
scalars.  Finally, supersymmetry seems to
be intimately connected with gravity.  String theory
is the only theory we know -- and quite possibly the
only theory of any kind -- which incorporates gravity
and gauge interactions.  But we have learned during the
past two years that what we call string theory is really
a code for some larger structure, whose precise nature
we only partly understand.  Supersymmetry seems
to play a fundamental role in this structure.  Moreover,
it is hard to see how we will be able to make any
sense of string theory if supersymmetry does not
survive to comparatively low energies.

So there are a number of arguments that suggest
that nature might be supersymmetric, and that supersymmetry
might manifest itself at energies of order the weak
interaction scale.  There is also some weak
but tantalizing experimental evidence, as I will
discuss in these lectures.  On the
other hand, while many find these
arguments for supersymmetry
persuasive, I believe one should be skeptical.  Perhaps
some sort of technicolor theory can ultimately
explain the presence of light scalars, or perhaps
our ideas about hierarchy and fine tuning are not
correct, there is no supersymmetry
and yet there is a light fundamental scalar
with mass less than a TeV.  Experiment
should decide this question over the next ten to fifteen years.

In these lectures, I will discuss several aspects of low
energy supersymmetry.  I will try, first, to make
the case that supersymmetry may underlie
the physics of weak interactions.  I hope, also, to
make clear that in the event that supersymmetry is
discovered, the pattern of soft breaking -- the masses
of the partners of ordinary fields -- should reveal
a great deal.  Indeed, one could imagine that the
high energy physics of the next century will
be devoted to unraveling this pattern, and deciphering
its meaning, in much the same way that studying
low energy weak interactions revealed the nature
of the underlying gauge theory.  

The rest of these lectures are organized as follows.
In the next section, some
features of $N=1$ supersymmetric theories
are briefly reviewed.
Then I discuss the hierarchy problem, and how
it may be resolved in a supersymmetric framework.
I explain why $N=1$ supersymmetry is special,
and introduce a supersymmetric version of the standard model
(the MSSM).
In section 3, I introduce the notion of soft breaking and count
the parameters of the model.  A simple ansatz for these
parameters is described, which is highly predictive, satisfies
constraints from rare processes,
and automatically yields $SU(2) \times U(1)$
breaking.   This is followed by a critique
of the ansatz, and a more careful examination of the
various phenomenological constraints
on the soft breaking parameters.
These constraints arise from direct searches, from flavor
changing neutral currents and other rare processes, and
from
the requirement of $SU(2) \times U(1)$ breaking.
I also discuss coupling constant unification and the
question of dark matter.

Sections 4-6 are devoted to the problem of
supersymmetry breaking.  Various mechanisms
for dynamical supersymmetry breaking are
introduced. I then turn to the question:  what
is the scale of supersymmetry breaking?
Two possibilities are considered:  a scale
intermediate between $M_Z$ and $M_p$,
and a much lower energy scale, of order $10's$
to 100's of TeV.  The virtues and difficulties
of both approaches are discussed, as well
as some dramatic experimental consequences.
In particular, low energy supersymmetry seems
a natural framework in which to understand
the $e^+ e^- \gamma \gamma$ event seen
by the CDF detector at Fermilab.

Finally, section 7 is devoted to some
questions in string theory.  I examine how
some of the issues raised earlier look in the context
of strings.  I also discuss how the problem
of vacuum stability/instability -- which
appears to be one of the most
fundamental  questions of string dynamics --
appears in a low energy framework.  I consider
this problem first  from the perspective of weakly
coupled strings, and then turn to the
strongly coupled limit (``M theory").
M theory may well turn out to be more
appropriate to the description of the real world then
weakly coupled string theory.

\section{N=1 Supersymmetry and ``Low Energy" Physics}

\subsection{What is N=1 Supersymmetry?}

In four dimensions, it is possible to have as many as
eight supersymmetries.  It is unlikely that theories with $N>1$
play any role in low energy physics.  First, such
theories are nonchiral.  Second, it is
virtually impossible to break supersymmetry in theories with $N>1$.
The symmetries simply prevent one from
writing any term in the effective lagrangian
which could yield supersymmetry breaking.

\smallskip
\noindent
{\bf Exercise:}  Check these statements, using
results from Lykken's lectures.
\smallskip

So if nature is supersymmetric at scales comparable
to the weak scale, there is almost certainly
only one supersymmetry.  The basic supersymmetry
algebra is then
\beq
\{ Q_{\alpha}, Q_{\dot \beta}^* \}
=2 \sigma_{\mu} P^{\mu}.
\label{nequalsone}
\eeq
There is a straightforward
recipe for constructing theories with this
symmetry.  The construction has been
described in Joe Lykken's lectures at this school,
and an excellent introduction is provided by
the text of Wess
and Bagger \cite{wessbagger}.
We will first consider the case of global supersymmetry;
later, we will consider the generalization to
local supersymmetry.   There are two irreducible
representations of the supersymmetry algebra containing
fields of spin less than or equal to one.  These
are the chiral and vector superfields.  Chiral fields
contain a Weyl spinor and a complex scalar; vector
fields contain a Weyl spinor and a (massless)
vector.
In superspace (using the conventions of \cite{wessbagger}),
a chiral superfield may be written as
\beq
\Phi(x,\theta)= A(x) + \sqrt{2}
\theta \psi(x) + \theta^2F + \dots
\label{chiralfield}
\eeq
Here $A$ is the complex scalar, $\psi$ the fermion,
and $F$ is an auxiliary field.  
Under a supersymmetry transformation with anticommuting
parameter $\zeta$, the component fields transform
as
\beq
\delta A= \sqrt{2} \zeta \psi
\label{atransform}
\eeq
\beq
\delta \psi = \sqrt{2} \zeta F + \sqrt{2} i
\sigma^{\mu} \bar \zeta \partial_{\mu} A~~~~~
\delta F= -\sqrt{2}i \partial_{\mu} \psi \sigma^{\mu} \bar
\zeta.
\label{psitransform}
\eeq

\smallskip
\noindent
{\bf Exercise}: For a chiral field, construct the supersymmetry
{\it operators.}  Verify that if $F$ has an expectation
value, supersymmetry is broken and that $\psi$ is
the Goldstone fermion (in particular, the supercurrent
contains a term $\langle F \rangle \sigma^{\mu} \psi$).
\smallskip

For vector superfields, the physical content is most
transparent in a particular gauge (really a class
of gauges) know as Wess-Zumino gauge.  This gauge
is  analogous to the Coulomb gauge in QED.  In that case,
the gauge choice breaks manifest Lorentz invariance,
but Lorentz invariance is still a property of physical amplitudes.
Similarly, the choice of Wess-Zumino gauge
breaks supersymmetry, but physical
quantities are still supersymmetric.  In this gauge, the
vector superfield may be written as
\beq
V=-\theta \sigma^{\mu} \bar \lambda
A_{\mu} + i \theta^2 \bar \theta \bar \lambda
-i \bar \theta^2 \theta \lambda + \half
\theta^2 \bar \theta^2 D.
\label{vectorfield}
\eeq
Here $A_{\mu}$ is the gauge field, $\lambda_{\alpha}$
is the gaugino, and $D$ is an auxiliary field.
The analog of the
gauge invariant field strength is a chiral field:
\beq
W_{\alpha} = -i \lambda_{\alpha}
+ \theta_{\alpha}D -\smallfrac i 2
(\sigma^{\mu} \bar \sigma^{\nu} \theta)_{\alpha} F_{\mu \nu}
+ \theta^2 \sigma^{\mu}_{\alpha \dot \beta} \partial_{\mu}
\bar \lambda^{\dot \beta}.
\label{wdefinition}
\eeq

To construct an action with $N=1$ supersymmetry,
one starts with a set of chiral superfields, $\Phi^{i}$,
transforming in various representations of some gauge
group ${\cal G}$.   For each gauge generator, there
is a vector superfield, $V^a$.  The most general
renormalizable lagrangian, written in superspace, is
\beq
{\cal L} = \sum_i \int d^4 \theta
\Phi_i^{\dagger} d^V \Phi_i + \sum_a
{1 \over 4 g_a^2} \int d^2 \theta W_{\alpha}^2
+c.c. + \int d^2 \theta W(\Phi_i) + c.c.
\label{superspacel}
\eeq
Here $W(\Phi)$ is a holomorphic function of chiral superfields
known as the superpotential.

In terms of the component fields, the lagrangian
takes the form (again in Wess-Zumino gauge):
\beq
{\cal L} = \sum_i \left ( \vert
D \phi_i \vert^2 + i\psi_i \sigma^{\mu} D_{\mu}
\psi_i^* + \vert F_i \vert^2 \right )
\label{componentla}
\eeq
$$
\quad \quad 
-\sum_a{1 \over 4 g_a^2}\left ( F_{\mu \nu}^{a2} -i \lambda^a \slash
D \lambda^{a*} -\half (D^a)^2 \right )$$
$${\quad \quad 
+ i \sqrt{2}\sum_{ia}
 g^a \psi_i T^a \lambda^a \phi^* + c.c.}$$
$${
+ \sum_{ij} {1\over 2}{\partial^2 W \over \partial \phi^i
\partial \phi^j} \psi^i \psi^j.}$$
Here I have already changed notation, and have used
$\phi_i$ to denote the scalar component of $\Phi$.
This is a common practice, and I will often use it where
it will not lead to (too much) confusion.
I have also solved for the auxiliary fields $F_i$ and
$D_a$ using their equations of motion:
\beq
F_i={\partial W \over\partial \phi_i}~~~~~
D^a = g^a \sum_i \phi_i^* T^a \phi_i.
\label{fcomponent}
\eeq
The first two lines are just the gauge invariant kinetic
terms for the various fields, as well as potential terms
for the scalars.  The third line corresponds to Yukawa
interactions of the gaugino and matter fields, with
strength controlled by the strength of the gauge
interactions.  The last line yields fermion
mass terms and Yukawa couplings
among the various chiral fields.

As you have already seen (for example in the lectures
of Greene and Peskin at this school) it is often useful
to consider effective lagrangians valid below
some energy scale.  In this case there is no
restriction of renormalizability.  On the other
hand, one usually does want to expand the
lagrangian in powers of momenta.  It is a simple
matter to generalize eq.~(\ref{superspacel}) to include
arbitrary terms with up to two derivatives:
\beq
{\cal L} = \sum_i
\int d^4 \theta K(\Phi^{i*} \Phi^i)
+ \sum_{ab}\int d^2 \theta
f_{ab}(\Phi) W_{\alpha}^a W^{\alpha b} +c.c.
+ \int d^2 \theta W(\Phi_i) +c.c.
\label{nonrenormalizable}
\eeq
The functions $W$ and $f$ are holomorphic
functions of the chiral fields (otherwise the last
two terms are not supersymmetric); $K$
is unrestricted.  It is not
hard to generalize the component lagrangian,
and I will leave that as an exercise.  Note that
among the couplings are now terms:
\beq
\fourth \rm Re~ f_{ab}(\phi_i)
F_{\mu \nu}^a F^{\mu \nu b} + \fourth
{\rm Im} ~f_{ab}(\phi_i)
F_{\mu \nu}^a \tilde F^{\mu \nu b} + {\partial f_{ab}
\over \partial \phi_i} F_i \lambda^a \lambda^b.
\label{sampleterms}
\eeq

\smallskip
\noindent
{\bf Exercise:}  Work out the terms in the effective
lagrangian.  Be more careful about factors of two
than I have been above.
\smallskip

\subsection{Why should we be Interested in Low Energy
Supersymmetry?}

\begin{figure}[htbp]
\centering
\centerline{\psfig{file=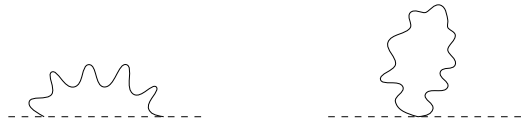,width=7cm,angle=-90}}
\caption{One loop corrections to the Higgs mass in the standard
model.}
\label{oneloophiggs}
\end{figure}

It is quite possible that supersymmetry will be discovered
in the near future.  In the fall of 1996, for example, LEP
will run at an energy of 172 GeV.  During 1997, it will
run at about 190 GeV.  So it is conceivable that by the time
these proceedings appear, superpartners of some
ordinary particles might have been seen.  Absent a discovery,
however, there are three classes of reasons for thinking
that supersymmetry might have something to do with
nature, and that it might be broken at a scale
comparable to the scale of weak interactions, rather than
at some enormous energy such as the Planck scale.  The first
of these is the ``hierarchy problem" \cite{wilson},
the fact
that light fundamental scalars don't seem to make
sense in quantum field theory.  Here light
means light compared to the largest interesting energy
scale.  One might imagine that this scale is the Planck
mass or unification scale.  To see the difficulty, consider the minimal standard
model, with a single Higgs doublet.  Then the diagrams of
fig.~\ref{oneloophiggs} are quadratically divergent in the ultraviolet,
so that the Higgs mass is given by a formula of the
form
\beq
m_H^2
= (m_H^2)_o + {\alpha_2 \over4\pi} \Lambda^2.
\label{quadraticdiv}
\eeq
If $\Lambda$ is some extremely large scale, the
Higgs mass can only be small if there is a delicate
balance between classical and quantum
efects.   Such a conspiracy
would represent a stupendous familure of
dimensional analysis.  In other
words, the natural value of the Higgs mass would seem to be
the largest scale in nature.  That the Higgs
boson is light (compared to, e.g., $M_p$) might
indicate that it has struture on a
scale comparable to the scale of weak interactions (technicolor).
But failures of dimensional analysis often reflect the
presence of symmetries.  The only known symmetry
which can suppress the quadratically divergent corrections
is supersymmetry.
 
\begin{figure}[htbp]
\centering
\centerline{\psfig{file=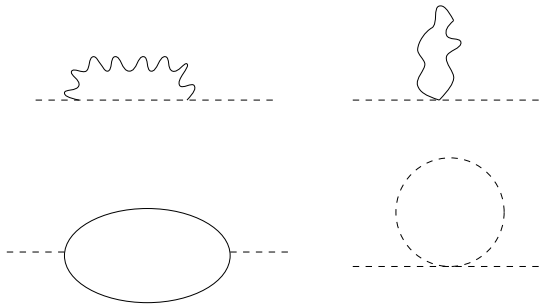,height=4cm,angle=-90}}
\caption{One loop gauge corrections to the scalar mass
in a simple model.
}
\label{oneloopgauge}
\end{figure}

To get some practice with supersymmetric theories,
let's check that in a simple model,
the quadratic divergences do indeed cancel.  Take
a $U(1)$ theory, with (massless)
chiral fields $\phi^+$
and $\phi^-$.  Before doing any computation, it is easy
to see that provided we work in a way
which preserves supersymmetry, there can be no
quadratic divergence.  In the limit that the mass term
vanishes, the theory has a chiral symmetry under which
$\phi^+$ and $\phi^-$ rotate by the same phase,
\beq
\phi^{\pm} \rightarrow e^{i \alpha}
\phi^{\pm}.
\label{chiralsymm}
\eeq
This symmetry forbids a mass term in the superpotential,
$\Lambda \phi^+ \phi^-$, the only way a supersymmetric mass
term could appear.  The actual diagrams we need
to compute are shown in
 fig.~\ref{oneloopgauge}.  Since we are only
interested in the mass, we can take the external momentum
to be zero.  It is convenient to choose Landau gauge for
the gauge boson.  In this gauge the gauge boson propagator is
\beq
D_{\mu \nu} = -i(g_{\mu \nu} -
{q_{\mu} q_{\nu} \over q^2}) {1 \over q^2}
\label{landaugauge}
\eeq
so the first diagram vanishes.  The second, third and fourth
are straightforward to work out from the basic lagrangian.
One finds:
\beq
I_b= g^2(i)(-i){3 \over (2\pi)^4} \int {d^4 k \over k^2}
\label{ibeqn}
\eeq
\beq
I_c= g^2(i)(-i){(\sqrt{2})^2 \over (2\pi)^4} \int {d^4 k \over
k^4 } {\rm tr} (k_{\mu} \sigma^{\mu} k_{\nu} \bar \sigma^{\nu})
\label{iceqn}
\eeq
\beq
\quad \quad =-{4g^2 \over (2 \pi)^4} \int {d^4 k \over k^2}
\label{ideqn}
\eeq
\beq
I_c= g^2(i)(-i){1 \over (2\pi)^4} \int {d^4 k \over k^2}.
\label{ieeqn}
\eeq
It is easy to see that the sum, $I_a +I_b+I_c+I_d=0.$

Clearly if nature is supersymmetric, supersymmetry is broken.
We want, then, to ask, what is the likely scale of supersymmetry
breaking?  We can modify our computation above so as to
address this question.  Suppose that supersymmetry breaking
induces a mass for the scalars, $\widetilde m^2$, but the fermions
remain massless.  Then only $I_d$ changes;
\beq
I_d \rightarrow {g^2 \over (2\pi)^4}
\int {d^4k \over k^2 - \widetilde m^2}
\label{idbroken}
\eeq
$$\quad \quad=-i{g^2 \over (2 \pi)^4} \int {d^4 k_E \over k_E^2 + \tilde
m^2}$$
$$=\quad \quad \rm \widetilde m^2~
{\rm independent} +
{ig^2 \over 16 \pi^2} \widetilde m^2 \ln(\Lambda^2 /
\widetilde m^2).$$
We have worked here in Minkowski space, and I have
indicated factors of $i$ to assist the reader in obtaining
the correct signs for the diagrams.
In the second line, we have performed a Wick rotation.  In the
third, we have separated off a mass-independent part,
since we know that this is cancelled by the other diagrams.

\smallskip
\smallskip
\noindent
{\bf Exercise:}  Verify the expressions for $I_a - I_d$.
\smallskip
\smallskip

Summarizing, the one loop mass shift is
\beq
\delta \widetilde m^2 =
-{g^2 \over 16 \pi^2} \widetilde m^2 \ln(\Lambda^2 /\widetilde m^2).
\label{oneloopshift}
\eeq
  Note that the mass shift  is proportional to $\widetilde m^2$,
the supersymmetry breaking mass, as we would expect
since supersymmetry is restored as $\widetilde m^2 \rightarrow 0$.
In the context of the standard model, we see that
the scale of supersymmetry breaking cannot be much larger
than the the Higgs mass scale itself.  Roughly speaking, it
can't be much larger than this scale by factors of order
$1/\sqrt{\alpha_W}$, i.e., factors of order $6$.

A second reason to suspect that low energy supersymmetry
might have something to do with nature comes from string
theory.  From the lectures at this school, you have
surely gained the sense that supersymmetry is
intrinsic to string theory.  It is natural
to suspect that any consistent
fundamental theory must be supersymmetric.

By itself, this is not enough to argue that supersymmetry
should survive to low energies.  But it has
been known for some time that there are a vast array
of supersymmetric {\it solutions} of string theory.
A general feature of these solutions is that if
supersymmetry is unbroken in some
lowest order approximation, the theory
remains supersymmetric to all orders.  Moreover,
non-supersymmetric solutions are problematic:  typically
the vacuum is unstable already at one loop.
While one cannot claim to have understood how string
dynamics might break supersymmetry and choose
some particular ground state, it is almost impossible
to imagine how a sensible ground state could emerge
in a non-supersymmetric vacuum.   So if string theory
describes nature, low energy supersymmetry is almost
certainly a prediction.

Finally, there are some small experimental hints
that supersymmetry might be true.  The most dramatic
of these is the unification of couplings \cite{couplingunity}.
To understand
this, we first introduce a supersymmetric extension
of the standard model, known as the ``Minimal Supersymmetric
Standard Model," or MSSM.  
In this model, the gauge symmetry is still taken to
be $SU(3) \times SU(2) \times U(1).$   Each gauge generator
is now associated with a vector multiplet, so there is a gaugino
for each gauge boson.  Similarly,
all of the known quarks and leptons
are promoted to chiral multiplets.  In other words,
each left-handed fermion of the standard model
now has a complex scalar partner with the same
quantum numbers.  Finally, instead of one Higgs
doublet, there must be two, each containing a boson
and a fermion.  Otherwise the model suffers from
anomalies.  We will denote these by
$H_U$ and $H_D$, with hypercharge $\pm 1$, respectively.

Knowing the representation content of the theory,
we can work out the $\beta$-functions of the
different groups.  The general expression for the one-loop
$\beta$-functions is
\beq
b_o= {11 \over 3} C_A - {2 \over 3} \sum_{i=1}^{n_f}
c_2^i - {1 \over 3} \sum_{j=1}^{n_{\phi}}
 c_2^j.
\label{bonon}
\eeq
Here the sum over $i$ runs over all of the left-handed
fermions of the theory, while that over $j$ runs
over the scalars.
For the gauginos, $c_2 = C_A$, so for a supersymmetric
theory with $n_f$ chiral
fields in the fundamental representation we obtain
\beq
b_o = 3 C_A - {1 \over 2} n_f.
\label{bosusy}
\eeq

For $SU(3)$, with three generations, this gives
$b_o=3$.  Similarly, for $SU(2)$, one obtains
$b_o=-1$ (remember to keep track of the two
Higgs doublets).  For the $U(1)$, some care is
required with the normalization of the charge.
Let us assume that the three gauge groups are
unified in $SU(5)$.  In that case, all of the generators
must be normalized in the same way.  In a singlet
generation, one has a $\bar 5$ and $10$.
The $\bar 5$ contains the $\bar d$ quark
and the lepton doublet.   For this representation,
the $SU(3)$ and $SU(2)$
generators satisfy
${\rm tr} T^2 = {1/2}.$  The corresponding $U(1)$ generator
is then
\beq
\widetilde Y = {\sqrt{3 \over 20}}~
{\rm diag}(2/3,2/3,2/3,-1,-1).
\label{hypercharge}
\eeq
In other words, $\widetilde Y$ is related to the
conventional hypercharge generator by:
\beq
\widetilde Y =
\sqrt{3/20} Y \quad \quad  g^{\prime} = g_5 \sqrt{3/5}
\label{hyperchargenorm}
\eeq
(remember that the hypercharge boson is taken
to couple to $Y/2$).    From this it follows
that $sin^2(\theta_w) = 3/8$.   

With this information, one can now run the gauge
couplings to high energy.  The simplest thing to do is to
assume that all of the new particles predicted by
supersymmetry have the same mass (say a few hundred
GeV up to a TeV).
At one loop, the couplings satify
\beq
\alpha_i^{-1}(M_Z)=\alpha_i^{-1}(M_{GUT})+ {b_o^{(i)}\over 4 \pi}
\ln(M_{GUT}/M_Z).
\label{couplingrunning}
\eeq
Including also two loop corrections,
one obtains quite good agreement. The couplings
fail to unify in nonsupersymmetric theories.
This might be a coincidence, but it is quite suggestive.\footnote{Since the
weak and electromagnetic
couplings are far better known than $\alpha_s$, one usually
computes $\alpha_s$, making some assumptions
about thresholds both at the GUT scale and at the
supersymmetric scale.  One actually finds that,
with the simplest assumptions for these, that one predicts
a value of $\alpha_s$ slightly too large.  This problem,
and possible solutions, have been discussed in \cite{damien}.}

\subsection{More on the MSSM}

Above, we have introduced the fields of the MSSM,
and explained their gauge quantum numbers.  Let us
develop this model further.
The lagrangian includes, first,
the gauge invariant kinetic terms for all of the
fields.  In a supersymmetric theory, this
includes Yukawa couplings of gauginos to matter
fields and quartic scalar couplings from the $D^2$ terms.
The usual Yukawa couplings of fermions arise from
terms in the superpotential.  These can
be written in the form
\beq
W= H_UQy_u \bar u + H_D Q y_D \bar d
+ H_D L y_L \bar e.
\label{yukawas}
\eeq
In this expression, the quark and lepton superfields
are understood to carry a flavor
index, and the $y$'s are $3 \times 3$ matrices.
If $H_U$ and $H_D$ have non-zero
masses, this gives mass for quarks and squarks, leptons
and sleptons.

\smallskip
\smallskip
\noindent
{\bf Exercise}:  Check that their are a set of supersymmetric
ground states with equal, non-zero expectation values for $H_u$
and $H_D$ i.e., that the energy vanishes if
$H_U=H_D =  \left({\st 0 \atop \st v}\right)$.  Check that
one obtains equal masses for bosons and fermions, first
for the quarks and leptons, then for the gauge bosons
and gauginos.
\smallskip
\smallskip

There are other terms which can also be present in the
superpotential.  These include the ``$\mu$-term,''
$\mu H_U H_D$.  This is a supersymmetric mass term
for the Higgs fields.  We will see later that we need
$\mu \sim M_Z$ to have a viable phenomenology.

A set of dimension four terms
which are permitted by the gauge symmetries raise
much more serious issues.
For example, one can have terms
\beq
\bar u_f \bar d_g \bar d_h \Gamma^{fgh}
+ Q_f L_g \bar d_H \lambda^{fgh}.
\label{bviolating}
\eeq
These couplings violate $B$ and $L$!  This is our first
serious setback.  In the standard model,
there is no such problem.  The leading operators permitted
by gauge invariance are four fermi operators of dimension
$6$, and it is easy to imagine that they are suppressed by
some very large mass scale.

If we are not going to simply give up, we need to suppress
$B$ and $L$ violation at the level
of dimension four terms.
This presumably requires additional
symmetries.  There are various possibilities
one can imagine.
\begin{enumerate}
\item Global continuous symmetries:
It is hard to see how such symmetries could
be preserved in any quantum theory of gravity,
and indeed in string theory, there is a theorem
which asserts that there are no global continuous
symmetries \cite{banksdixon}.
\item   Discrete symmetries:  Discrete symmetries can
be gauge symmetries, and indeed such symmetries are
common in string theory.  These symmetries
are often ``R symmetries," symmetries which
do not commute with supersymmetry.
\end{enumerate}

A simple (though not unique), solution to the problem
of baryon and lepton number violation by dimension
four operators is known as $R$-parity or ``matter parity."
Under this symmetry, all
ordinary particles are even, while their superparters
are odd.  Imposing this symmetry
immediately eliminates all of the dangerous operators.
For example,
\beq
\int d^2 \theta \bar u \bar d \bar d
\sim \psi_{\bar u} \psi_{\bar d} \,\tilde{\!\bar d}
\label{bexample}
\eeq
(we have changed notation again:  the tilde here
indicates the superpartner of the ordinary field,
i.e., the squark).  This operator is clearly odd under the
symmetry.

More formally, we can define this symmetry as
the transformation on superfields:
\beq
\theta_{\alpha} \rightarrow - \theta_{\alpha}
\label{rparity}
\eeq
\beq
(Q_f, \bar u_f \bar d_f, L_f \bar e_f) \rightarrow
-(Q_f, \bar u_f \bar d_f, L_f \bar e_f) 
\label{rparitya}
\eeq
\beq
(H_U ,H_D) \rightarrow (H_U H_D).
\label{rparityb}
\eeq

While imposing this symmetry solves the immediate
problem, it also has a striking consequence:  the lightest
of the new particles predicted by supersymmetry (the LSP)
is {\it stable}.  In order to avoid various cosmological
disasters, this particle must be electrically neutral.
It is then, inevitably, very weakly interacting.  This
in turn means:
\begin{itemize}
\item  The generic signature of R-parity conserving
supersymmetric theories is events with missing energy.
\item  Supersymmetry is likely to produce
an interesting dark matter candidate.
\end{itemize}

\noindent
In most of what follows, we will assume a conserved
$R$-parity.

\subsection{LSP as the Dark Matter}

A stable particle is not necessarily a good dark matter
candidate.  But we can make a crude calculation
which indicates that the LSP density is in a suitable
range to be the dark matter.  Consider particles, $X$,
with mass of order $100$ GeV interacting
with weak interaction strength.  Their
annihilation cross sections go as
$G_F^2 E^2$.  So,
in the early universe, the corresponding
interaction rate is of order
\beq
\Gamma \approx \rho_X
G_F^2 E^2 \approx \rho_X G_F^2 T^2.
\label{interactionrate}
\eeq
These interactions drop out of equilibrium when
\beq
\Gamma \sim H \sim {T^2 /M_p},
\label{equilibcond}
\eeq
i.e., when
\beq
\rho_X \sim {G_F^2 \over M_p}\sim 10^{-9}.
\label{densityest}
\eeq
$T$ here is of order 1~GeV, so
\beq
{\rho_X \over \rho_{\gamma}} \sim 10^{-9}.
\label{rhoxrhog}
\eeq
This means that the $X$ particles have a number density
today similar to that of baryons.  So if their masses are of order
$100$ GeV, their density can be of order the closure density.
This estimate is quite crude, but more careful
studies indicate that the LSP can be the dark matter
for a broad range of parameters.

So while it is disturbing that we need to impose additional
symmetries in order to avoid proton decay, it is also
exciting that this leads to a possible solution of one of the
most critical problems of cosmology:  the identity of
 the
dark matter.

\section{A First Look at Supersymmetry Breaking}

\subsection{Explicit Soft Breaking of Supersymmetry}

If supersymmetry is a symmetry of nature, it is almost
certainly an exact, local symmetry.
This is the case, for example, in string theory.   Even if
string theory is not the correct underlying
theory, it is hard to imagine how supersymmetry
might arise ``by accident."  So supersymmetry
must be spontaneously broken.
We will devote a great deal of attention in these
lectures to the problem of spontaneous breaking of supersymmetry.
However, it turns out that most schemes for spontaneous
breaking yield an effective lagrangian,
at low energies, which is supersymmetric except for
explicit, soft supersymmetry violating
terms.  So we will begin by simply adding
such terms to the effective lagrangian of the MSSM,
without altering the dimensionless couplings.

The addition of such soft terms is compatible with our
original hope that supersymmetry could solve the hierarchy
problem.  Indeed, precisely because these terms are
soft, they induce at most logarithmic ultraviolet divergences.
As an example, consider a theory with a single
massless chiral superfield, $\Phi$, with superpotential
\beq
W= {\lambda \over 3} \Phi^3.
\label{wesszumino}
\eeq
To the supersymmetric lagrangian for this theory, we add
an explicit mass term for the scalar component of
$\phi$, i.e., we take the lagrangian
to be
\beq
{\cal L} = \vert \partial_{\mu} \phi \vert^2
+ i \psi \slash \partial \psi^* + \lambda \psi \psi \phi + c.c.
-\vert \lambda \vert^2 \vert \phi \vert^4 -m_{sb}^2 \vert \phi
\vert^2.
\label{introsoft}
\eeq

\begin{figure}[htbp]
\centering
\centerline{\psfig{file=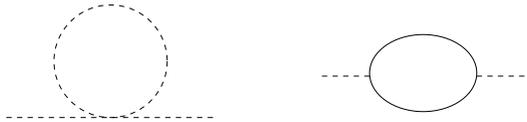,width=7cm,angle=-90}}
\caption{One loop corrections to scalar masses arising
from Yukawa couplings.
}
\label{oneloopyukawas}
\end{figure}

At one loop, the scalar masses receive corrections from the diagrams
of fig.~\ref{oneloopyukawas}.   One has
\beq
I_a = 4 {\lambda^2 \over (2\pi)^4}
\int {d^4 k \over k^2 -m_{sb}^2}
\label{softcorrections}
\eeq
$${I_b = (-)2 {\lambda^2 \over (2\pi)^4}
\int {d^4 k \over k^2 }{\rm tr}{1}.}$$
Here the trace is over the Weyl indices and gives a factor
of two, so as expected the two diagrams cancel in the
supersymmetric limit.

\smallskip
\smallskip
\noindent{\bf Exercise:}  Check the signs and combinatorics
of these diagrams.
\smallskip
\smallskip

For large momenta,
\beq
{1 \over k^2 - m_{sb}^2} \approx
{1 \over k^2} + {m_{sb}^2 \over k^4}.
\label{largekapprox}
\eeq
The leading divergence cancels, and one obtains for
the mass shift,
\beq
\delta m^2 \approx -{\vert \lambda \vert^2 \over 4 \pi^2}
m_{sb}^2 \ln (\Lambda^2 /m^2).
\label{finalshift}
\eeq
Note the negative sign.  This will be important when
we come to consider the breaking of $SU(2) \times U(1).$

It is not difficult to make a list of the symmetry breaking
interactions which are soft in this sense \cite{allowedsoft}. These include:
\begin{enumerate}
\item  Masses for scalars (either of the form $\phi^* \phi$,
as above, or $\phi \phi$ when permitted by symmetries).
\item  Masses for gauginos.
\item  Cubic couplings for scalars, of the form
$\phi_i\phi_j\phi_k +c.c.$, but not $\phi_i^* \phi_j \phi_k$.
\end{enumerate}

We can understand this list in a simple way, by considering
{\it spontaneous} supersymmetry breaking.  The auxiliary fields,
$F_i$ and $D^a$, which sit in the chiral and vector supermultiplets,
are candidate order parameters for this breaking.  To see this,
consider the commutation relations (dropping the indices $i$
and $a$, for simplicity):
\beq
\{Q_{\alpha}, \psi_{\beta} \}= \sqrt{2}
F \epsilon_{\alpha \beta} + \dots\ .
\label{fbreaking}
\eeq
\beq
\{Q_{\alpha}, \lambda_{\beta} \}= \sqrt{2}
D \epsilon_{\alpha \beta} + \dots\ .
\label{dbreaking}
\eeq
So if $F$ or $D$ have expectaion
values, the $Q_{\alpha}$'s do not annhilate
the vacuum and supersymmetry is broken.
In general, these fields could be elementary or composite,
but for our present discussion, we will assume that they
are elementary.  

Now suppose we add to the MSSM some fields $\Phi$ and
$V$ with nonvanishing auxiliary components, and
which are neutral under the standard model gauge group.
Then the soft breaking terms
a-c can arise through terms in the effective action such as:
\beq
{1 \over M^2} \int d^4 \theta \Phi^{\dagger}\Phi
Q^{\dagger} Q ={\vert \langle F \rangle \vert^2 \over M^2}
\widetilde Q^{\dagger} \widetilde Q + (\rm derivative ~\rm terms).
\label{softa}
\eeq
\beq
{1 \over M}\int d^2 \theta \Phi W_{\alpha}^2
W^{\alpha a}= {\langle F \rangle \over M} \lambda^a \lambda^a
+ \dots
\label{softb}
\eeq
\beq
{1 \over M}\int d^2 \theta \Phi H Q \overline U
= {\langle F \rangle \over M} H \widetilde Q \tilde{\bar u} + \dots\ .
\label{softc}
\eeq

Without a microscopic theory of supersymmetry breaking,
all of the soft terms are independent.
It is interesting to ask, in the MSSM, how many soft breaking
parameters are there?  More precisely, let's count the parameters
of the model beyond those of the minimal standard model
with a single Higgs doublet. Having imposed R parity,
the number of Yukawa couplings is the same in both
theories, as is the number of gauge couplings and $\theta$
parameters.  The quartic couplings of the Higgs fields
are completely determined in terms of the gauge couplings.
So the ``new" terms arise from the soft
breaking terms, as well as the $\mu$ term for the Higgs fields.
We will speak loosely of all of this as the ``soft breaking"
lagrangian.  Suppressing flavor indices:
\beq
{\cal L}_{sb}=\widetilde Q^* m_Q^2 \widetilde Q+
\tilde{\bar u}^* m_{\bar u}^2 \tilde{\bar u}+
\,\tilde{\!\bar d}^* m_{\bar d}^2 \,\tilde{\!\bar d}
\label{softbreakingl}
\eeq
$$
~~~~+
 \tilde L^* m_L^2 \tilde L
+ \tilde{\bar e}^* m_{\bar e}^2 \tilde{\bar e}
$$
$$
~~~~~~+
H_U \widetilde Q A_u \tilde{\bar u} +
H_D \widetilde Q A_d \,\tilde{\!\bar d} +H_D L
A_l \bar e + \rm c.c.
$$
$$
~~~~~~+m_i \lambda \lambda +c.c.
+m_{H_U}^2 \vert H_U\vert^2+m_{H_D}^2 \vert H_D \vert^2
+ \mu B H_U H_D + \mu \psi_H \psi_H.
$$
The matrices $m_Q^2$, $m_{\bar u}^2$, and so on
are $3\times 3$ hermitian matrices, so they
have nine independent entries.  The matrices $A_u$, $A_d$,
etc., are general $3 \times 3$ complex matrices,
so they each possess $18$ independent entries.
Each of the gaugino masses is a complex number,
so
these introduce $6$ additional parameters.  The
quantities $\mu$ and $B$ are also complex; this
is four more.  In total, then, there
are $111$ new parameters.    As in the standard model,
not all of these parameters are real; we are free to
make field redefinitions.  The counting, however,
is significantly simplified if we just ask how many
parameters there are beyond the usual $17$ of the
minimal theory, since this counting uses up
most of our freedom.

To understand what redefinitions are possible beyond
the transformations on the quarks and leptons which
go into defining the usual KM parameters, we need
to ask what are the symmetries of the MSSM before
introducing the soft breaking terms and the
$\mu$ term (the $\mu$ term is more or less
on the same footing as the soft breaking terms,
since it is of the same order of magnitude; as we will
discuss later, it might well
arise from the physics of supersymmetry breaking).
Apart from
the usual baryon and lepton numbers, there are two
more.  The first is a Peccei-Quinn symmetry, under which
which two Higgs superfields rotate by the same phase,
while the right handed quarks and leptons rotate by
the opposite phase.  The second is perhaps more
interesting.  It is an ``R" symmetry.   By definition,
and $R$ symmetry is a symmetry of the Hamiltonian
which does not commute with the supersymmetry generators.
Such symmetries can be continuous or discrete.
In the case of continuous $R$-symmetries, by convention,
we can take the $\theta$'s to transform by
a phase $e^{i \alpha}$.  Then the general transformation law
takes the form
\beq
\lambda_i \rightarrow e^{i \alpha} \lambda_i
\label{rgauginos}
\eeq
for the gauginos, while for the elements of a chiral multiplet
\beq
\Phi_i(x,\theta) \rightarrow e^{i r_i\alpha}\Phi(x,\theta e^{i \alpha}),
\label{rchiral}
\eeq
or, in terms of the component fields,
\beq
\phi_i \rightarrow e^{i r_i\alpha}\phi_i~~~~~
\psi_i \rightarrow e^{i (r_i-1)\alpha}\psi_i~~~~~
F_i \rightarrow e^{i (r_i-2)\alpha}F_i\ .
\label{rcomponents}
\eeq
In order that the lagrangian exhibit a continuous $R$ symmetry,
the total $R$ charge of all terms in the superpotential must
be two.
In the MSSM, we can take $r_i=2/3$ for all of the
chiral fields.

The soft breaking terms, in general, break two of the three
lepton number symmetries, the $R$ symmetry and
the Peccei-Quinn symmetry.  So there are four non-trivial
field redefinitions which we can perform.  In addition,
the minimal standard model has two Higgs parameters.
So from our $111$ parameters, we can subtract a total
of six, leaving $105$ as the number of {\it new} parameters
in the MSSM.

Clearly we would like to have a theory which predicts these
parameters.  Later, we will study some candidates.
To get started, however, it is helpful to make
an ansatz.  The simplest thing to do is suppose that
all of the scalar masses are the same, all of the
gaugino masses the same, and so on.  It is necessary
to specify also a scale at which this ansatz holds,
since it will not be respected by renormalization to lower
energies.  Almost all investigations of superymmetry
phenomenology assume such a degeneracy at a large
energy scale, typically the reduced Planck mass,
$M=M_p/\sqrt{8\pi}$.
This assumption is sometimes considered part of the
definition of the MSSM, but I will use MSSM simply
to refer to the particle content of the model.   It is often
said that degeneracy is automatic in supergravity models,
so this is frequently called the supergravity (``SUGRA")
model, but as well will see, supergravity by itself makes
{\it no} prediction of degeneracy.  In any case, the ansatz
consists of the statement that at the high energy scale:
\begin{enumerate}
\item  All of the scalar masses are the same,
$\widetilde m^2 = m_o^2$.  This asssumption is called
``universality" of scalar masses.
\item  The gaugino masses are the same
$M_i = M_o$.  This is referred to as the
``GUT" relation, since it holds in simple
grand unified models.
\item  The soft-breaking cubic terms are assumed to be given by
\beq
{\cal L}_{tri}= A(H_U Q y_u \bar u + H_D Q y_d \bar d
+ H_D L y_l \bar e).
\label{sbcall}
\eeq
The matrices $y_u$,$y_d$, etc. are
the same matrices which appear in the Yukawa
couplings.  This is the assumption of
``proportionality."
\end{enumerate}

Note that with this ansatz, if we ignore possible
phases, five parameters are required to specify
the model ($m_o^2,M_o,A,B\mu,\mu$).  One of these
can be traded for $M_Z$, so this is quite an
improvement in predictive power.  In addition,
this ansatz automatically satisfies all constraints
from rare processes.   With
this assumption, these constraints are automatically
satisfied.  On the other
hand, we will want to ask: just how plausible are these
assumptions?   We will try to address this
question later in these lectures.

\subsection{$SU(2) \times U(1)$ Breaking}

In the MSSM, there are a number of general statements
which can be made about the breaking of $SU(2) \times U(1)$.
The only quartic couplings of the Higgs fields arise
from the $SU(2)$ and $U(1)$ $D^2$ terms.  The general
form of the soft breaking mass terms has been described
above.  So, before worrying about any detailed ansatz
for the soft breakings, the Higgs potential is
given, quite generally, by
$$
V_{\it Higgs} = m_{H_U}^2  \vert H_U \vert^2
+m_{H_D}^2\vert H_D \vert^2 -m_3^2(H_U H_D+h.c.)$$
\beq
~~~~~~
+ {1 \over 8}(g^2 + g^{\prime 2})(\vert H_U \vert^2
-\vert H_D \vert^2)^2 + {1 \over 2}
g^2 \vert H_U H_D \vert^2.
\label{vhiggsa}
\eeq
This potential by itself conserves CP;
a simple field redefinition removes any
phase in $m_{12}^2$.  (As we will discuss shortly,
there are many other possible souces of CP violation
in the MSSM.)
The physical states in the Higgs sector are usually
described by assuming that CP is a good symmetry.
In that case, there are two $CP$-even scalars,
$H^o$ and $h^o$, where by convention, $h^o$
is the lighter of the two.  There is a $CP$-odd neutral
scalar, $A^o$, and charged scalars, $H^{\pm}$.
At tree level, one also defines
a parameter,
\beq
{\rm tan} (\beta ) = {\vert \langle H_U \rangle \vert
\over \vert \langle H_D \rangle \vert} \equiv {v_1/v_2}.
\label{tanbeta}
\eeq
Note that with this definition, as $\tan(\beta)$
grows, so does the Yukawa coupling of the $b$-quark.

To obtain a suitable vacuum, there are two constraints
which the soft breakings must satisfy: 
\begin{enumerate}
\item  Without the soft breaking terms,
$H_U=H_D$ ($v_1=v_2=v$) makes the $SU(2)$
and $U(1)$ D terms vanish, i.e., there is no quartic
coupling in this direction.  So the
energy is unbounded below unless
\beq
m_{H_U}^2 + m_{H_D}^2-2\vert m_3 \vert^2
>0.
\label{conditiona}
\eeq
\item  In order to obtain symmetry
breaking, the Higgs mass matrix must
have a negative eigenvalue.  This gives
the requirement:
\beq
\vert m_{3}^2 \vert^2 > m_{H_U}^2 m_{H_D}^2.
\label{conditionb}
\eeq
\end{enumerate}

When these conditions are satisfied, it is straightforward
to minimize the potential and determine the spectrum.
One finds that
\beq
m_A^2 = {m_{12}^2 \over \sin (\beta)
\cos (\beta)}.
\label{masssquared}
\eeq
It is conventional to take $m_A^2$ as one of the parameters.
Then one finds that the charged Higgs masses are given by
\beq
m_{H^{\pm}}^2=m_W^2 + m_A^2,
\label{chargedhiggs}
\eeq
while the neutral Higgs masses are:
\beq
m^2_{H^o,h^o}={1 \over 2} \left
(m_A^2+m_Z^2\pm\sqrt{(m_A^2+m_Z^2)^2-4m_Z^2
m_A^2\cos (2\beta)} \right ).
\label{neutralhiggs}
\eeq

\smallskip
\smallskip
\noindent
{\bf Exercise:}  Derive these formulae.
\smallskip
\smallskip

Note the inequalities:
\beq
m_{h^o} \le m_A
\label{inequalities}
\eeq
$$m_{h^o} \le m_Z$$
$$m_{H^{\pm}} \ge m_W.$$

\begin{figure}[htbp]
\centering
\centerline{\psfig{file=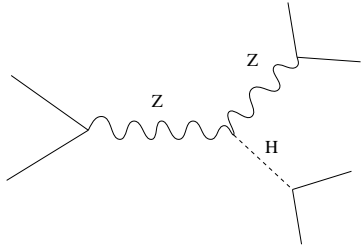,height=3.25cm,angle=-90}}
\caption{Leading contribution to Higgs production in
$e^+~e^-$ annihilation.
}
\label{higgsproduction}
\end{figure}

The middle relation is particularly interesting since LEP II
will have enough energy to find $h^o$
if $m_{h^o} \le 95$ GeV.  Thus it would appear
that it might be able to rule out (or discover) the
MSSM at this machine.
The basic process
through which one hopes to discover the higgs
uses the $Z-Z-h$ vertex, as in fig.~\ref{higgsproduction}.
From observations on the $Z$ pole, one already
has a limit of about 60 GeV.  However, these are tree level relations.  We will
turn shortly to the issue of radiative corrections,
and will see that these can be quite substantial --
LEP II will not be able to rule out the MSSM \cite{topquarkloops}.

There are other states in the MSSM which are likely
discovery channels for supersymmetry. Particularly
important among these are the ``charginos,"
linear combinations of the partners of the
$W^{\pm}$ and $H^{\pm}$.  The mass matrix for these states,
denoted $w^{\pm}$ and $\tilde h^{\pm}$
is given by
\beq
{\cal M}_{\chi^{\pm}}=
\left ( \matrix{ M_2 & gv_1 \cr gv_2 & \mu} \right ).
\label{charginomass}
\eeq
There are also four neutral fermions, referred to as
the neutralinos, $w^o$, $b$, $\tilde h_U^o$, $\tilde h_D^o.$
The lightest of these states is a natural dark matter candidate.

\subsection{Why is one Higgs mass negative?}

Within the simple ansatz, there is a natural way to understand
why $m_{H_U}^2 <0$ while $m_{H_D}^2 >0$ \cite{ibanezross,agcw}.
What is special
about $H_U$ is that it has an ${\cal O}(1)$ coupling
to the top quark.  (If $\tan(\beta)$ is very large, of order
40--50, $H_D$ has a comparable coupling
to the $b$ quark).  We saw earlier in the Wess Zumino
model that at one loop, there is a negative
renormalization of the soft breaking scalar masses.
This calculation can be translated to the MSSM,
with a modification for the color and $SU(2)$ factors.
One obtains:
\beq
m_{H_U}^2 = (m_{H_U})_o^2 - {6y_t^2 \over 16
\pi^2}
\ln (\Lambda^2/m^2) \widetilde m_t^2.
\label{mhushift}
\eeq
\beq
\widetilde m_{t}^2 = (\widetilde m_{t})_o^2 -
{4 y_t^2\over 16 \pi^2}
\ln (\Lambda^2/m^2) \widetilde m_H^2.
\label{huoneloop}
\eeq

So we see that loop corrections involving the top quark Yukawa
coupling reduce both the Higgs and the stop masses, but
the reduction is larger for the Higgs. If $\Lambda \sim M_p$,
and the typical soft breakings are of order a TeV, these
corrections are ${\cal O}(1)$, so one needs a full
renormalization group analysis to determine
if $SU(2) \times U(1)$ is broken.  For this we need the full
set of renormalization group equations.  These can be derived
along the lines of the calculations we have already presented
for the gauge and Yukawa contributions to the soft mass
renormalizations.  If only $y_t$ is large,
they can be written rather compactly:
\vbox{
\beq
\mu {\partial \over \partial \mu} m_{H_U}^2
= {1 \over 8 \pi^2} [3y_t^2(\mhu\! + m_{\bar t}^2\! + m_{Q_3}^2
+\vert A_U^{33} \vert^2)]
-{1 \over 2 \pi^2} \left[\smallfrac34 \vert M_2 \vert^2 g_2^2 + 
\smallfrac14  \vert M_1 \vert^2 g_1^2\right]
\label{rgea}
\eeq
\beq
\mu {\partial \over \partial \mu}m_{\bar t}^2
= {1 \over 8 \pi^2}[2y_t^2(\mhu\! + m_{\bar t}^2\! + m_{Q_3}^2
+\vert A_U^{33} \vert^2)]
-{1 \over 2 \pi^2} \!\left[ \smallfrac43 \vert M_3 \vert^2 g_3^2 +
\smallfrac49 \vert M_1 \vert^2 g_1^2\right]
\label{rgeb}
\eeq
$$
\mu {\partial \over \partial \mu}m_{Q_3}^2 =
{1 \over 8 \pi^2} \!\left[2y_t^2(\mhu + m_{\bar t}^2 + m_{Q_3}^2
+\vert A_U^{33} \vert^2)\right]
$$
\beq
~~~~~~~~~~~~~~~~~~
 -{1 \over 2 \pi^2} \left[\smallfrac43 \vert M_3 \vert^2 g_3^2 +
\smallfrac34 \vert M_2 \vert^2g_2^2+ \smallfrac1{36}
\vert M_1 \vert^2 g_1^2\right].
\label{rgec}
\eeq }

For scalars besides $H_U$ and the third generation squarks
one has only the contribution from diagrams involving
intermediate gauginos:
\beq
\mu {\partial \over \partial \mu}m_i^2=
-{1 \over 2\pi^2}\sum_a g_a^2 \vert M_a \vert^2 c_{ai}
\label{rgeq}
\eeq
where $c_{ai}$ denotes the appropriate Casimir ($1/2$ for
particles in the fundamental representation)
while gaugino masses satisfy:
\beq
\mu {\partial \over \partial \mu} \left({M_a^2 \over
g_a^2} \right)=0.
\label{rgee}
\eeq

It is straightforward to integrate these equations numerically.
For a significant range of parameters, one does obtain
suitable breaking of $SU(2) \times U(1)$.

\subsection{Radiative Corrections to the Higgs Mass Limit}

At tree level, the form of the Higgs potential
is highly constrained.  The quartic terms are
exactly known.  Once supersymmetry is broken,
however, there can be corrections to the quartic
terms from radiative corrections.\footnote{In sufficiently
complicated models, there can be tree level
corrections to the quartic couplings.  This does
not occur in the MSSM, but it can occur in models
with singlets.}  These corrections are soft, in that
the susy-violating four point functions vanish
rapidly at momenta above the supersymmetry breaking
scale.  Still, they are important in determining
the low energy properties of the theory, such as the
Higgs vev's and the spectrum.

\begin{figure}[ht]
\centering
\centerline{\psfig{file=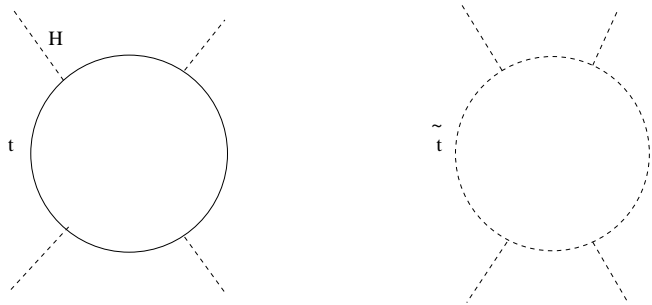,height=4cm,angle=-90}}
\caption{Corrections to quartic Higgs couplings
from top loops.
}
\label{topcorrections}
\end{figure}

The largest effect of this kind comes from
loops involving top quarks \cite{topquarkloops}.
It is not hard to get
a rough estimate of the effect.  Consider the diagrams
of fig.~\ref{topcorrections}, and suppose that $\widetilde m_t \gg m_t.$
In this limit, we can omit the top squark from the
computation.  The result will be logarithmically
divergent, and we can take the cutoff to be
$\widetilde m_t.$  So we have
\beq
\delta \lambda = (-1)y_t^4 \times 3 \int
{d^4k \over (2 \pi)^4}{\rm Tr} {1 \over (\slash k -m_t)^4}
\label{deltalambda}
\eeq
\beq
~~~~=-{12 i y_t^4 \over 16\pi^2}
\ln(\widetilde m_t^2 /m_t^2).
\label{resultforlambda}
\eeq
The implications of this result are left for the
following exercise.

\smallskip
\smallskip
\noindent
{\bf Exercise:}  Verify these formulae.  Evaluate $y_t$
in terms of $m_t$ and $\sin (\beta)$.  Show that
to this level of accuracy,
$$m_h^2 < m_Z^2 \cos(2 \beta) + {12 g^2 \over 16 \pi^2}
{m_t^4 \over m_W^2} \ln(\widetilde m^2  m_t^2).$$
Check that you are in rough agreement with the
numerical results in ref.~\cite{topquarkloops}.

\subsection{Constraints on Soft Breakings}

While we have stressed that there is a large number of
soft breaking parameters, there are also many experimental
constraints.  These come from the failure of direct searches
to see superpartners of ordinary fields, and also
from indirect effects.

\begin{figure}[ht]
\centering
\centerline{\psfig{file=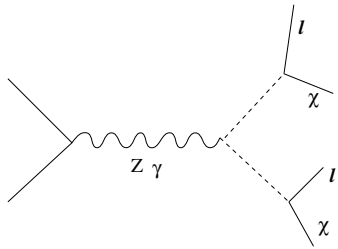,height=3.25cm,angle=-90}}
\caption{Slepton production in $e^+~e^-$ annihilation.
}
\label{sleptonpro}
\end{figure}

The direct searches are easy to describe, and
production and decay rates can be computed given
knowledge of the spectrum, since the couplings
of the fields are known.  If $R$ parity is conserved, the LSP is
stable and
 weakly interacting, so the characteristic signal
for supersymmetry is {\it missing energy.}  For example,
in $e^+ e^-$ colliders, one can produce slepton pairs,
if they are light enough, through the diagram of fig.~\ref{sleptonpro}.
These then decay, typically, to a lepton and a neutralino,
as indicated.  So the final state contains a pair of acoplanar
leptons, and missing energy.  From such processes,
one has limits of order $45$ GeV for all of the charged
states of the MSSM.  LEPII will eventually raise most
of these limits to a range of order 95 GeV.  Chargino masses are limited
in this way to about 65 GeV; early reports from this winter's
run indicate that the chargino limits are approaching 85 GeV.

One can obtain stronger limits -- and in many cases greater
discovery potential -- from hadron machines.  For example,
because they are strongly coupled and they are octets
of color,
gluinos have very substantial production cross sections
in hadron collisions.  They can be produced both by
$q \bar q$ and $gg$ annihilation.  Gluinos can
decay to a large number of channels, and many of these are used in
setting limits on the gluino mass.  These limits range
from $150$ to $225$ GeV, depending on the model
assumptions (e.g., R-parity violating or not, masses
of squarks) used in the analysis.  Similar limits apply to
squarks.  Limits on charginos are similar
to those currently set by LEP.

Rare processes provide another set of strong constraints
on the soft breaking parameters.
In the simple ansatz all of the scalar masses are
the same at some very high energy scale.  However,
if this is true at one scale, it is not true
at all scales, i.e., these relations are {\it renormalized}.
Indeed, all $105$ parameters are truly parameters,
and it is not obvious that the assumptions of universality
and proportionality are {\it natural.}  On the other
hand, there are strong experimental constraints which
suggest some degree of degeneracy.

\begin{figure}[ht]
\centering
\centerline{\psfig{file=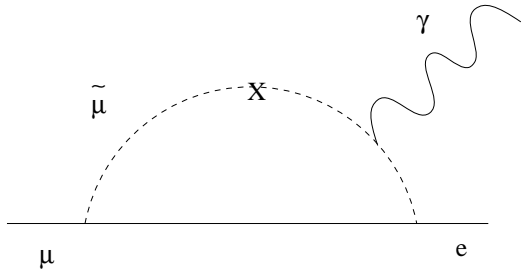,height=3.5cm,angle=-90}}
\caption{Contribution to $\mu \rightarrow e \gamma$.
}
\label{fig:muegamma}
\end{figure}

As one example, there is no reason, a priori, why the mass
matrix for
the $\tilde L$'s (the partners
of the lepton doublets)  should be diagonal in the
same basis as the charged leptons.  If it is not,
there is no conservation of separate lepton numbers,
and the decay $\mu \rightarrow e \gamma$
will occur (fig.~\ref{fig:muegamma}).  To see that we are potentially
in serious trouble, we
can make a crude estimate.
If we suppose that the characteristic
scale of the supersymmetric particles is
of order $m_W$, then the branching
ratio goes as
\beq
BR \sim \left({\alpha_2 \over 4 \pi}\right)^2
\left( {\delta \widetilde m_{L}^2 \over m_{\rm susy}^2}\right)^2
\label{muegamma}
\eeq
where $m_{\rm susy}^2$ denotes a typical
susy-breaking mass scale, and $\delta \widetilde m_{L}^2 $
denotes some off-diagonal term in the mass matrix.
This branching ratio is about $10^{-4}$ if
the off diagonal terms are large, 
so we are off by about $6$ orders of magnitude.
in this case.  Things get better as the susy-breaking
scale grows as $m_{\rm susy}^2$, but even if this
scale is $500$ GeV, we have to explain
a dimensionless number of order $10^{-2}$.  With the assumption
that $m_L^2$ is proportional to the unit matrix at some
large scale, the separate lepton numbers are exact at that
scale, and, of course, there is no problem.

\begin{figure}[ht]
\centering
\centerline{\psfig{file=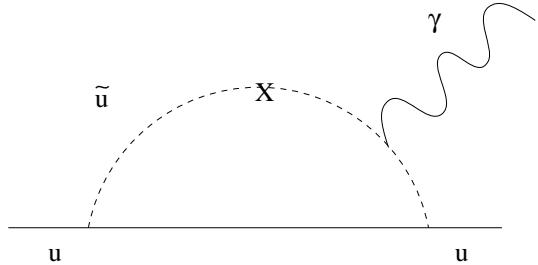,height=3.5cm,angle=-90}}
\caption{Contribution to $d_n$ in supersymmetric theories.
}
\label{fig:dn}
\end{figure}

Another troublesome constraint arises from the neutron electric
dipole moment, $d_n$.  Any non-zero value of this quantity
signifies $CP$-violation.  Currently, one has
$d_n \le 10^{-25} e ~{\rm cm}.$  The soft breaking terms
in the MSSM contain many new sources of $CP$-violation.
Even with the assumptions of universality and proportionality,
the gaugino mass and the $A$, $\mu$ and $B$ parameters
all are complex, and can violate $CP$.  At the quark level,
the issue is that one loop diagrams can generate
a quark dipole moment, as in fig.~\ref{fig:dn}.  Note that this
particular diagram is proportional to the phases of the
gluino and the $A$ parameter.  It is easy to see that even
if $m_{\rm susy} \sim 500 \rm GeV$, these phases must be
smaller than about $10^{-2}$.  More detailed estimates
can be found in \cite{dncalc}.

\smallskip
\smallskip
\noindent
{\bf Exercise:}  Make this estimate.
\smallskip
\smallskip

$CP$ is violated in the real world, so it is puzzling
that all of the soft supersymmetry violating
terms should preserve $CP$ to such a high degree.
In fact, it is usually said that one of the triumphs of the
minimal standard model is that it explains the observed
CP violation with a CP-violating phase of order one.
It is thus a serious challenge to understand
why $CP$ should be such a good symmetry if
nature is supersymmetric.  Some possible explanations
will be considered later in this lecture.

\begin{figure}[htbp]
\centering
\centerline{\psfig{file=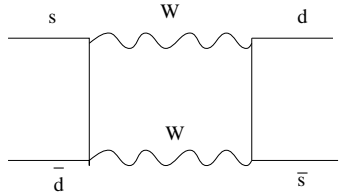,height=2.5cm,angle=-90}}
\caption{Contribution to $K \leftrightarrow \overline K$
in the standard model.
}
\label{wbox}
\end{figure}

So far, we have discussed constraints on the slepton
degeneracy and CP-violating phases.  There are
also constraints on the squark masses arising from
various flavor violating processes.  In the standard
model, the most famous of these are strangeness
changing processes, such as $K \overline K$ mixing.
One of the early triumphs of the standard model
was that it successfully explained why this mixing
is so small.  Indeed, the standard model gives a quite good estimate
for the mixing.   This was originally used to
predict -- amazingly accurately -- the
charmed quark mass \cite{gaillardlee}.%
\footnote{The computation of $K$--$\overline K$ mixing
is discussed in many texts and reviews.  See, for
example, \cite{georgi,nirreview}.}
The mixing recieves
contributions from box diagrams such as the one shown in fig.~\ref{wbox}.
If we consider first, only the first two generations
and ignore the quark masses (compared to $M_W$),
we have that
\beq
{\cal M}(K^o \rightarrow \overline K^o)
\propto (V_{di} V^{\dagger}_{is})(V_{sj}^{\dagger}V_{jd}) =0.
\label{leadingkkbar}
\eeq
Including fermion masses leads to terms
in ${\cal L}_{\it eff}$ of order
\beq
{\alpha_W \over 4 \pi} {m_c^2 \over M_W^2}
G_F \ln(m_c^2/m_u^2) (\bar s \gamma^{\mu}
\gamma_5 d) (\bar d \gamma^{\mu} \gamma^5 s)
+ \dots
\label{kkbaroperator}
\eeq
The matrix element of the operator appearing here can be estimated
in various ways, and one finds that this
expression roughly saturates the observed value.  Similarly,
the CP-violating part (the ``$\epsilon$" parameter) is in
rough accord with observation, for reasonable values of the
KM parameter $\delta$.

\begin{figure}[htbp]
\centering
\centerline{\psfig{file=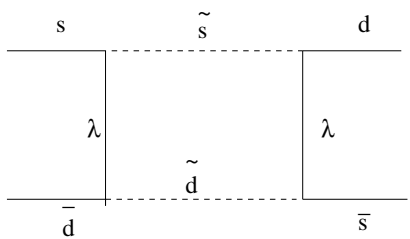,height=3cm,angle=-90}}
\caption{Gluino exchange contribution to kaon
mixing in the MSSM.
}
\label{gluinobox}
\end{figure}

In supersymmetric theories, if squarks are degenerate,
there are similar cancellations.  However, if they are
not, there are new, very dangerous contributions.
The most serious is that indicated in fig.~\ref{gluinobox}, arising
from exchange of gluinos and squarks.  This is nominally larger
than the standard model contribution
by a factor of $({\alpha_s/ \alpha_W})^2\approx 10$.
Also, the standard model contribution vanishes in the
chiral limit, whereas the gluino exchange does not,
and this leads to an addtional enhancement of
nearly an order of magnitude.  On the other hand, the
diagram is highly suppressed in the limit of
exact universality and proportionality.  Proportionality
means that the $A$ terms in eq. (\ref{softbreakingl})
are suppressed by factors of light quark masses,
while universality means that the squark propagator,
$<\tilde q^* \tilde q>$,
is proportional to the unit matrix in flavor space.
So there are no appreciable off-diagonal terms
which can contribute to the diagram.
On the other hand, there is surely some degree
of non-degeneracy.  One finds that even if the
characteristic susy scale is 500~GeV,
one needs degeneracy in the down squark sector
at the part in $10^2$ level.
More generally, it vanishes in the limit that the
squark mass matrix is diagonal in the same basis as the
quark mass matrix~\cite{nirseiberg}.

So
$K$--$\overline K$ mixing tightly constrains the down squark
mass matrix.  The imaginary part provides further
constraints.  There are
also strong limits on $D$--$\overline D$ mixing, which
significantly restrict the mass matrix in the up squark
sector.
Other important constraints on soft breakings come
from other rare processes.  The current
situation is carefully surveyed in ref.~\cite{masiero}.

\begin{figure}[htb]
\centering
\centerline{\psfig{file=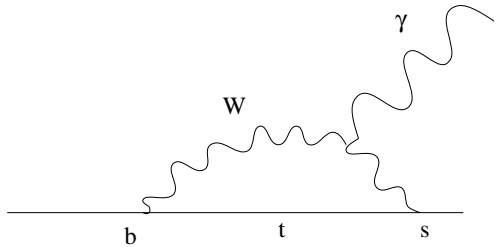,height=3.25cm,angle=-90}}
\caption{Standard model contribution to $b \rightarrow s+ \gamma$.
}
\label{bsgamma}
\end{figure}

Violations of universality might be expected to
be largest in the third generation, so
a quite powerful set of constraints comes from the
process $b \rightarrow s \gamma$. This has been measured
by CLEO (see Drell's talk at this school).  One finds
\beq
BR(B\rightarrow s \gamma) = (2.32
\pm 0.51 \pm 0.29 \pm 0.32) \times 10^{-4}.
\label{bsgrate}
\eeq
This is quite consistent with the standard model prediction, which
arises from the $W$ loop in fig.~\ref{bsgamma}:
\beq
BR(B\rightarrow s \gamma)= (3.25\pm 0.30 \pm 0.40) \times 10^{-4}.
\label{bsgsm}
\eeq
In the MSSM -- and for that matter, in any theory with
two Higgs doublets -- there is at least one additional contribution,
coming from a loop with the charged $W$ replaced by
a charged Higgs \cite{chargedhiggs}. Unless there are cancellations,
the mass of the charged Higgs must be greater than about 300~GeV.
This is
troubling.  After all, we would like the Higgs mass
parameters to be numbers comparable to $M_Z$.  
One might worry that such massive Higgs imply a fine tuning
of parameters at the $10\%$ level or worse.  Whether this is
acceptable or not, only time will tell.  
In many models, it turns out that the charged
Higgs must be quite massive for other reasons as well.

\begin{figure}[ht]
\centering
\centerline{\psfig{file=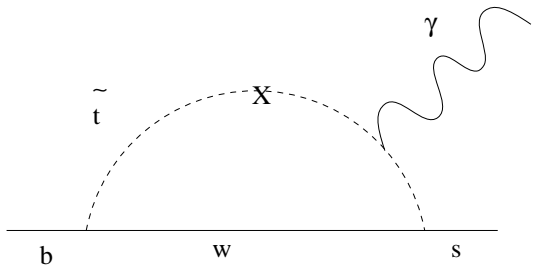,height=3.5cm,angle=-90}}
\caption{Additional contributions to $b \rightarrow s+ \gamma$
in the MSSM.
}
\label{susybsg}
\end{figure}

Under some circumstances, there
{\it are} additional, negative corrections to the rate which
can ameliorate this problem.
The
principal new contributions in the MSSM are diagrams
with stops and charginos in the loop (fig.~\ref{susybsg}.) \cite{bg}.
These are unimportant
unless these particles are relatively light.  In models
with exact degeneracy at the high scale, such light
stops are not implausible in view of the renormalization
group equations for the stop mass, which tend to lower
the stop mass.  So this is perhaps some support for
the notion of high scale universality (though as noted
before, detailed phenomenological studies still often
produce large values for the charged Higgs mass).

It has been suggested that light stops and charginos
might be of interest for another reason.  At the time
that these lectures were presented, there was a substantial
discrepancy between the standard model value of
\beq
R_b = {\Gamma(Z \rightarrow b \bar b) \over
\Gamma(Z\rightarrow
X)}.
\label{rbdef}
\eeq
This discrepancy, averaging data from LEP
and SLAC (SLD) was about $3.4 \sigma$.  Since
then, the ALEPH experiment has performed a further
analysis, obtaining a result in good agreement with the
standard model prediction
The world average now
differs from the standard model by about $2 \sigma$.
My own prejudice, which is widely shared, is that
this discrepancy is not real.  However, this is merely
prejudice, and it is interesting to note that the MSSM
with relatively light stops could account for most or
all of this discrepancy \cite{theoryrb}.
This is a situation which bears
further watching.

\subsection{The Rarest Process:  Proton Decay}

We can summarize our discussion of rare
processes up to now in the language of effective actions.
One can think of the action of the standard model
as an effective
action, obtained by integrating out some as yet
unknown physics at energy scales $M$
above the $W$ and $Z$ masses.
One of the virtues of this model is that several classes
of rare processes are automatically suppressed by factors
of $1/M$.  So as long as $M$ is large
enough, the model is compatible with experiment.
The problem in supersymmetry (and in any other framework,
like technicolor, which seeks to explain the scale of weak
interactions) is that  $M$ can't be much larger
than $M_Z$.  This means that other suppression mechanisms,
such as approximate or exact symmetries, must be found.
In the case of supersymmetry, reasonably simple
mechanisms do exist in each case.  Whether nature takes
advantage of them is another matter, of course.

We have already mentioned the most dramatic of
these constraints:  proton decay.  At the level of the
supersymmetric effective action, the gauge symmetry
permits dimension four terms in the superpotential
which violate $B$ and $L$.  If one integrates out the
squarks and sleptons, these generate various $B$ and
$L$ violating terms in the standard model with
coefficients of order $1/m_s^2$.  So clearly
these terms must be highly suppressed.  Up to now,
we have assumed that this suppression was achieved
through $R$ parity.  However, it is not really necessary
to eliminate every $B$ and $L$ violating operator in
order to insure proton stability.  For example, if $L$
is violated, but not $B$, the proton will be stable.
There are
many other constraints which must be considered,
such as $n - \bar n$ oscillations, and flavor changing
processes \cite{rparityviol}.
Still, it is possible that some of these couplings are not
too small.  This would significantly alter the problem
of susy detection.  Most importantly, the LSP would
not be stable, so missing energy would not necessarily
be an important signal.  One can easily imagine
that in string theory one might have some more intricate
discrete symmetry than the conventional $R$ parity,
which forbids some but not all of the
$R$-parity violating operators.

\smallskip
\smallskip
 \noindent
{\bf Exercise}:  Show how various $B$-violating four fermi
operators are generated by squark and slepton exchange,
starting with the general set of $B$ and $L$ violating
terms in the superpotential.  
\smallskip
\smallskip

Finally even in models with $R$ parity, the MSSM
possesses $B$ and $L$-violating dimension five operators
which are permitted by all symmetries \cite{weinbergfive}.
For example,
$R$-parity doesn't forbid such operators as
\beq
{\cal O}_5^a={1 \over M} \int d^2 \theta \bar u
\bar u \bar d e^+ \quad \quad {\cal O}_5^b =
{1 \over M} \int d^2 \theta QQQL.
\label{dimensionfive}
\eeq
These are still potentially  very dangerous.  When one integrates
out the squarks and gauginos, they will lead to
dimension six $B$ and $L$-violating operators in the
standard model with coefficients (optimistically) of order
\beq
{\alpha \over 4 \pi}{1 \over M m_{\rm susy}}.
\label{dimfivesix}
\eeq
Comparing with the usual minimal $SU(5)$ prediction,
and supposing that $M \sim 10^{16}$ GeV,
one sees that one needs a suppression of order $10^{9}$
or so.

Fortunately, such a suppression is quite plausible, at least
in the framework of supersymmetric guts \cite{dr}.
In a simple
$SU(5)$ model, for example, the operators of
eq.~(\ref{dimensionfive})
will
be generated by  exchange of the color triplet partners
of ordinary Higgs fields, and thus one gets
two factors of Yukawa couplings.  Also, in order that the
operators be $SU(3)$-invariant, the color indices must
be completely antisymmetrized, so more than one generation
must be involved.  This suggests further suppression by factors
of order Cabibo angles.   These numbers are typically of order
$10^{-9}-10^{-11}$.  More detailed studies of this question can be found
in the literature, and proton decay can be used to restrict
the parameter space of particular models.  But what is quite
striking is that we are automatically in the right range
to be compatible with experimental constraints, and perhaps
even to see something.  It is not obvious that things had to
be this way.

So far we have phrased this discussion in
terms of baryon-violating physics at $M_{GUT}$.
But whatever the underlying theory at $M_p$
may be, there is no reason to think that it should
preserve baryon number.  So one expects
that already at scales just below $M_p$, these
terms are present.  This would certainly be the case
in string theory.  If their coefficients were simply of order
$1/M_p$, the proton decay rate would be enormous.
Is there any reason to expect further suppression?   I believe
the answer is yes.  After all, the rate from Higgs
exchange in GUT's is so small because the Yukawa
couplings are small.  We do not really know why
Yukawa couplings are small, but it is natural to
suspect that this is a consequence of (approximate)
symmetries.
These same symmetries, if present would also
suppress dimension five operators from
Planck scale sources, presumably by a comparable
amount.

\section{The Origin of Supersymmetry Breaking}

\subsection{Simple Models of Supersymmetry Breakdown}

So far, we have treated supersymmetry as if it is explicitly
broken.  However, we have argued earlier that
supersymmetry must be an exact symmetry which
is spontaneously broken.  Fortunately, this is not incompatible with
the phenomenology
we have done up to now.
We have seen that soft breakings of the desired
type can arise in a theory with spontaneous breaking,
through operators in the effective lagrangian like
\beq
{1 \over M^2} \int d^4 \theta Z^{\dagger} Z
Q^{\dagger} Q \approx {\vert \langle F_Z \rangle \vert^2
\over M^2} \widetilde Q^{\dagger} \widetilde Q + \dots\ .
\label{typicalsoft}
\eeq

There are clearly a number of reasons to investigate the 
question of supersymmetry breaking.  First, we have seen
that without ad hoc assumptions, the MSSM has a huge number
of free parameters.   A theory of supersymmetry breaking
might make predictions for these quantities.  Another reason
concerns the hierarchy problem.  We have discussed the fact
that supersymmetry can eliminate the problem of quadratic
divergences.  But supersymmetry also has the
potential to {\it explain} the hierarchy \cite{wittendsb}.
In particular, if
supersymmetry is unbroken to lowest order in perturbation
theory in some theory, it is unbroken to all orders (with one
possible
exception, which we will discuss later).  This means that
supersymmetry breaking, if it does occur, must be smaller
than any power of the coupling, e.g.,
\beq
m_{\rm susy} = e^{-{8 \pi^2 \over Ng^2}}.
\label{hierarchies}
\eeq
It has been known for some time that
dynamical breaking can occur in four dimensions \cite{ads}.

Before turning to dynamical supersymmetry breaking,
it is instructive to consider models with
supersymmetry breaking at tree level.
We have seen that supersymmetry breaking is signalled
by a non-zero expectation value of an $F$ component
of a chiral or $D$ component of a  vector superfield.
Models involving
only chiral fields with no supersymmetric ground
state are called
to as ``O'Raifeartaigh'' models.  A simple example has three
singlet fields, $A,B,$ and $X$, with superspotential:
\beq
W= \lambda_1 A(X^2 - \mu^2) + \lambda_2 B X^2.
\label{oraif}
\eeq
With this superpotential, the equations
\beq
F_A={\partial W \over \partial A}=0 \quad \quad
F_B={\partial W \over \partial B}=0
\label{susybreaking}
\eeq
are incompatible.  To actually determine the expectation values
and the
vacuum energy, it is necessary to minimize the potential.
I'll leave this as an exercise.  Note, however,
that at this level not all of the fields are fully determined,
since the equation
\beq
{\partial W \over \partial X}=0
\label{xequation}
\eeq
can be satisfied provided
\beq
\lambda_1 A + \lambda_2 B =0.
\label{pseudoflat}
\eeq
This vacuum degeneracy is accidental, and as we will later
see, is lifted by quantum corrections.

It is also possible to generate an expectation value for a
$D$ term.  In the case of a $U(1)$ gauge symmetry
a term
\beq
\mu^2 \int d^4 \theta ~V = \mu^2 D
\label{dterm}
\eeq
is gauge invariant.  This is known as a ``Fayet-Iliopoulos
D term" and can lead to supersymmetry breaking.
For  example, if one has two charged fields, $\Phi^{\pm}$,
with charges $\pm 1$, and superpotential
$m \Phi^+\Phi^-$, one cannot simultaneously
make the two auxiliary $F$ fields and the
auxiliary $D$ field vanish.

\smallskip
\smallskip
\noindent
{\bf Exercise}:  Study the potentials in both
models and verify these statements.
\smallskip
\smallskip

One important feature of both types of models is that
at tree level, in the context of global
supersymmetry, the spectra are never realistic.
These spectra satisfy a sum rule,
\beq
\sum (-1)^F m^2  =0.
\label{treesum}
\eeq
Here $(-1)^F=1$ for bosons and $-1$ for fermions.
This guarantees that there are always light
states, and often color and/or electromagnetism
are broken.  These statements are not true of radiative
corrections, and of supergravity, as we will explain later.

It is instructive to prove this sum rule.  Consider
a theory of chiral fields only (no gauge interactions).
The potential is given by
\beq
V =\sum_i \left \vert {\partial W \over \partial \phi_i}
\right \vert^2.
\label{globalpotential}
\eeq
The boson mass matrix has terms of the form $\phi_i^* \phi_j$
and $\phi_i \phi_j + {\rm c.c.}$  The latter terms, as we will
now see, are connected with supersymmetry breaking.
The various terms in the mass matrix can be obtained by
differentiating the potential:
\beq
m_{i \bar j}^2 = {\partial^2 V \over \partial \phi_i
\partial \phi_{\bar j}^*} = {\partial^2 W \over \partial \phi_i
\partial \phi_k}{\partial^2 W^* \over \partial \phi_{\bar k}^* 
\partial \phi_{\bar j}^*},
\label{massa}
\eeq
\beq
m^2_{ij} ={\partial^2 V \over \partial \phi_i \partial \phi_j}
= {\partial W \over \partial \phi_k^*}
{\partial^3 W \over \partial \phi_k \partial \phi_i \partial \phi_j}.
\label{massb}
\eeq

The first of theses terms has just the structure of the
square of the fermion mass matrix,
\beq
{\cal M_F}_{ij} = {\partial^2 W \over 
\partial \phi_i \partial \phi_j}.
\label{massfermion}
\eeq
So writing the boson mass, ${\cal M}_B^2$
matrix on the basis $(\phi_i \; \phi_{\bar j}^*)$, we see that eq.~%
(\ref{treesum}) holds.

The theorem is true whenever a theory can
be described by a renormaliable effective
action.   We have seen that various
non-renormalizable terms in the effective
action can give additional contributions
to the mass, and with a little
thought it is clear that these will violate the
tree level sum rule.  Such terms arise
in renormalizable theories when one
integrates
out heavy fields to obtain an effective action
at some scale.  In the context of supergravity,
such terms are present already at tree level.
This is perhaps not suprising, given that these
theories are non-renormalizable and must
be viewed as effective theories from the
very beginning (perhaps the effective
low energy description of string theory).
Shortly, we will
discuss the construction of realistic models.
First, however, we turn to the issue of non-renormalization
theorems and dynamical
supersymmetry breaking.

\subsection{Non-Renormalization Theorems}

Non-supersymmetric theories have the property that
they tend to be generic; any term permitted by symmetries in the theory will appear in the effective action, with an order of magnitude
determined by dimensional analysis.\footnote{Possibly up
to a few powers of coupling.}
Supersymmetric theories are special in that this is not the
case.   This figures heavily in the duality story, as you
have heard at this school.  In $N=1$ theories, there are
non-renormalization theorems governing the superpotential
and the gauge coupling functions, $f$, of
eq.~(\ref{nonrenormalizable}).
These theorems assert that the superpotential is not renormalized
beyond its tree level value, while $f$ is at most renormalized
at one
loop \cite{nrtheorems}.$~$\footnote{There is an important
subtlety connected with these theorems.  Both
should be interpreted as applying only to a ``Wilsonian"
effective action, in which one integrates out physics above
some scale, $\mu$.  If infrared physics is included, the theorems
do not necessarily hold.  This is particularly important
for the gauge couplings.  In ref.~\cite{sv}, the
connection of the conventional $\beta$-function
and the Wilsonian one is explained in some detail.}

Originally, these theorems were proven by detailed study
of Feynman diagrams \cite{nrtheorems}.
Seiberg has pointed out that they
can be understood in a much simpler way \cite{seibergnr}.
Both the superpotential
and the functions $f$ are holomorphic functions of the chiral
fields, i.e., they are functions of the $\phi_i$'s and not
the $\phi_i^*$'s.  This is evident from their construction.
But the coupling constants of a theory may also
be thought of as {\it expectation values} of chiral
fields.  For example, consider a theory of a single chiral
field, $\Phi$,
with superpotential
\beq
W= \int d^2 \theta(m \Phi^2+ \lambda \Phi^3).
\label{wwz}
\eeq
We can think of $\lambda$ and $m$ as
expectation values of chiral fields, $\lambda(x,\theta)$
and $m(x,\theta)$.

\smallskip
\smallskip
\noindent
{\bf Exercise}:
One can make this more concrete, for example, by
treating these as massive fields. Replace
the term $m \Phi^2$ with $M
({1\over 2}S^2 -S m +S \Phi^2 )$.  Check that
$S$ has mass $M$ while $\Phi$ has mass $m$.
If $M$ is large,
it makes sense to think of the expectation value as frozen.
\smallskip
\smallskip

The lagrangian with $\lambda$ and $m$ has certain
symmetries.  In particular, if we first set $\lambda$
to zero, it has an $R$ symmetry under which $\Phi$
has $R$ charge one.  $\lambda$ has R charge $-1$
under this symmetry.  Now consider corrections to the
effective action.  For example, renormalizations of $\lambda$
in the superpotential necessarily involve positive powers
of $\lambda$.  But such terms (apart from $\lambda^1$)
have the wrong $R$ charge to preserve the symmetry.
So there can be no renormalization of this coupling.
There can be wave function renoralization, since the symmetries
allow the Kahler potential to
depend on $\lambda$, $K=K(\lambda^{\dagger} \lambda)$.

There are many interesting generalizations of these ideas,
and I won't survey them here, but I will mention two
further examples.  First, gauge couplings can clearly
be thought of in the same way, i.e., we can think of
$g^{-2}$ as a chiral field.  The real part of the scalar field
in this multiplet
couples to $F_{\mu \nu}^2$, but the imaginary part, $a$,
couples to $F \tilde F$.  $F \tilde F$
is a total derivative, and in perturbation theory there is
a symmetry under constant shifts of $a$.
But this means that the effective action should respect
this symmetry.  Because the gauge coupling function, $f$,
is holomorphic, this implies that
\beq
f(g^2)= \fourth g^{-2} + {\rm const}\ .
\label{onelooponly}
\eeq
The first term is just the tree level term.  The constant
term corresponds to one loop corrections.  There are
no higher order corrections in perturbation theory!
This is quite a striking result.  It is also paradoxical, since
the two loop $\beta$-functions for supersymmetric
Yang-Mills theories have been computed long ago,
and are in general non-zero.

Before explaining the resolution of this paradox,
there is one more non-renormalization
theorem which we can prove rather trivially here \cite{dtermnr}.
This is the statement
that if there is no Fayet-Iliopoulos $D$-term at
tree level, this term can be generated at most
at one loop.  To prove this, write the $D$ term as
\beq
\int d^4 \theta d(g,\lambda) V.
\label{dtermf}
\eeq
Here $d(g,\lambda)$ is some unknown function
of the gauge and other
couplings in the theory.  But if we think of $g$ and
$\lambda$ as chiral fields, this expression is only
gauge invariant if $d$ is a constant, corresponding to
a possible one loop contribution.  Such contributions
do arise in string theory.

In string theory, all of the parameters {\it are} expectation
values of
chiral fields.  Indeed, non-renormalization theorems
in string theory, both for world sheet\cite{newissues}\
and string perturbation theory\cite{dsnr}, were proved
by the sort of reasoning we have used above, long ago.

Returning to the paradox we raised
above, there is an important subtlety which must be
discussed here.   In textbooks, one often sees discussion
of something called the effective action, which is defined
to be the sum of one particle irreducible graphs.
The effective action we are describing here is something
different -- and significantly more meaningful -- called
the Wilsonian action.  This is defined as the action
obtained by integrating out physics above some energy
scale.   So, for example, it includes one particle reducible
diagrams containing massive fields, but it does not include
the low energy parts of loop graphs containing light
fields.  Diagrams of the first type give effective, local
interactions -- an example is $W$ and $Z$ exchange
at low energies in the standard model, which give rise
to the four fermi interaction.  Diagrams of the second type
give non-local interactions.  Since the non-renormalization
theorems certainly rely on locality, they need only
hold for the Wilsonian action.  The paradox of the
two loop renormalization of the gauge coupling is
resolved in just this way; there {\it is no}
renormalization beyond one loop for the
Wilsonian action \cite{sv}. Unfortunately, we do
not have a regulator for supersymmetric theories
analogous to lattice regulators in (non-chiral)
gauge theories, so it
is sometimes difficult to make this discussion concrete.
This problem will arise later, when we will want to discuss the
problem of unification in theories which are strongly coupled.

\subsection{Examples of Dynamical Supersymmetry Breaking}

In speaking of dynamical supersymmetry breaking, there
are two classes of models which one must consider.
\begin{enumerate}
\item  Models with moduli (e.g., generic string
vacuua).   In such theories, there is a continuous set
of degenerate vacuum states, at the classical
level, and the question is:  what
effects may lift the degeneracy?  We will see that
the degeneracy {\it is} often lifted, but
that there is usually no vacuum state at weak coupling.
\item  Models without flat directions.  In such
theories, the generic behavior is that
supersymmetry is unbroken, but under special circumstances,
supersymmetry is broken.
\end{enumerate}

These points can be illustrated by a theory known as
supersymmetric QCD.  By this I mean a theory with
$SU(N)$ gauge group, with $N_f$ flavors of ``quarks."
If the quarks are all massless, there is a large
classical vacuum degeneracy.
There are three distinct cases to consider:
\begin{enumerate}
\item  $N_f>N_c$:  The moduli space
is exact quantum mechically.
\item  $N_f=N_c$:  There is still an exact moduli space,
but the quantum moduli space is different than the
classical moduli space.
\item  $N_f<N_c$:   Non-perturbatively
a superpotential is generated in the effective
theory describing the moduli,
which lifts the flat directions.  However, in the regime
where one can do reliable calculations, the system
has no ground state.
\end{enumerate}

These statements can be understood, almost completely, on
symmetry grounds.  If there is no (classical) superpotential
for the quark and antiquark fields, 
$Q$ and $\overline Q$, the symmetry of the model at the
classical level is:
$SU(N_f)_L \times SU(N_f)_R \times U(1)_B
\times U(1)_A \times U(1)_R^{\prime}.$
$U(1)_B$ is just baryon number; $U(1)_A$ is the
analog of the familiar axial $U(1)$ of QCD.  
Quantum mechanically, a linear combination
of $U(1)_A$ and $U(1)_R^{\prime}$ is unbroken
by anomalies.
If we think of the gauge coupling
as a parameter, the anomaly in $U(1)_R^{\prime}$
can be cancelled by a shift of the field $a$.
Denote the quarks by $Q$ and $\overline Q$.  Then under
$SU(N_f)_L \times SU(N_f)_R \times U(1) \times U(1)_R\times
U(1)_R^{\prime}$, the charges of these fields are:
\beq
Q=\left(N,1,{N-N_f \over N_f},0\right)\quad
\quad\overline Q=\left(N,-1,{N-N_f \over N_f},0\right).
\label{qcharges}
\eeq
It is convenient to
 describe the rotation of the gauge coupling in terms
of the $\Lambda$ parameter of the theory,
$\Lambda= e^{-{8 \pi^2 \over b_o g^2}}$.  $\Lambda$
has charges
\beq
\Lambda=(0,0,0,0,{2(N-N_f) \over 3N -N_f}).
\label{lambdacharges}
\eeq

\smallskip
\smallskip
\noindent
{\bf Exercise:}  Verify that symmetries $U(1)_R$ and
$U(1)_R^{\prime}$ are non-anomalous.  It is necessary
to carefully work out the coupling of $a$ to $F \tilde F$,
and to check that under suitable shifts of $a$,
the shift in this term precisely cancels the usual anomaly
from fermion loops.
\smallskip
\smallskip

The classical moduli spaces of these models are easily
understood.  The scalar potential arises just from
the $D^2$ terms.  We can define matrix-valued fields,
\beq
D_i^{~j}= Q_i^* Q^j -\overline Q_i \overline Q^{*j} - {\rm tr~ terms}.
\label{matrixvalued}
\eeq
These vanish for
\beq
Q = \overline Q= {\rm diag}(a_1, \dots, a_{N_f}).
\label{flatdir}
\eeq
For $N_f<N_c$, this is the most general solution up to
symmetry transformations.  For $N_f \ge N_c$, there
are more general solutions.  For example, one can have
\beq
Q= {\rm diag}(v,v,v \dots ,v) \quad \quad \widetilde Q= {\rm diag}(v^{\prime},v^{\prime}
v^{\prime}\dots v^{\prime}).
\label{baryondirection}
\eeq

It is natural to
try to parameterize these flat directions in a gauge-invariant
way.  We can define a set of ``meson" operators,
\beq
M_a^b= \overline Q_aQ^b,
\label{mesonops}
\eeq
for any value of $N_f$; $a$ and $b$ are
flavor indices.  For $N_f \ge N_c$, we
can also define baryon and antibaryon operators,
\beq
B = Q^{a_1} \dots Q^{a_N} \quad \quad
\bar B = \overline Q^{a_1} \dots \overline Q^{a_N}.
\label{baryons}
\eeq
Note that the number of such operators depends
on $N_f$.  If $N_f=N_c$, there is one baryon and one
antibaryon, if $N_f=N_c+1$, there are $N_f$ baryons
and $N_f$ antibaryons, and so on.  

For the case $N_f = N_c-1$, let's check that the
$M$'s account for all of the light degrees of freedom.
For a generic point in the 
flat directions, the gauge group is broken
to $SU(N-N_f)$ (where $SU(1)$ is understood
to be trivial).  So the  $N^2-1$ gauge fields gain
mass.  When a gauge field gains mass, it ``eats"
an entire chiral multiplet.  So of the original
$2\times N \times N_f= N^2-N$ chiral fields, $N_f^2$ remain
massless.  There is precisely the number of $M$'s.
You can check that this counting works for
the other cases.

\smallskip
\smallskip
\noindent
{\bf Exercise:} 
\begin{enumerate}
\item   Verify that the $D$ term has the form of
eq.~(\ref{matrixvalued}), and that for $N_f<N_c$,
this is the most general solution.
\item  Check that the meson and baryon superfields
account for all of the light fields in the general
flat direction for $N_f \ge N$.
\end{enumerate}
\smallskip
\smallskip

Let's continue to focus on the case that $N_f \ge N-1$.
If the expectation values are large compared
to the scale of the theory, $\Lambda$, all of the
gauge bosons have masses large compared to
$\Lambda$.  The theory is then weakly
coupled, and a semiclassical analysis
should be valid.  The low energy theory consists
only of the fields $M$ and $B$.  The effective
action which describes these light fields should be
supersymmetric.  If supersymmetry is broken,
this should be understood as {\it spontaneous} breaking
in the low energy theory.  This can only occur
if the effective theory
contains a superpotential or a Fayet-Iliopoulos
term, but the latter possibility is ruled out
by our non-renormalization theorem (note
that this theorem was non-perturbative).  The form of any
possible superpotential, however, is greatly restricted
by the symmetries.  In order that the action
be invariant under $SU(N_f) \times SU(N_f)$,
the superpotential must be a function of
${\rm det} M$.  This determinant, however, vanishes
in the flat direction if $N_f > N$, and supersymmetry
cannot be broken.
If $N_f <N$, a superpotential is allowed, and the
symmetries uniquely determine its form.
In particular, noting that $W$ must have $R$ charge
two, and the $R$ charges of the fields in eqs.~(\ref{qcharges}) and
(\ref{lambdacharges}),
the superpotential must be
\beq
W = \left(\det \overline Q Q\right)^{-1 \over N-N_f}
\Lambda^{3N-N_f \over N-N_f}.
\label{wnp}
\eeq
This superpotential, in fact, respects all of the
symmetries for any value of $N_f<N$.  Note that
it also respects $U(1)_R^{\prime}$.

For $N_f=N-1$, we have noted that the theory is
weakly coupled, and that a semiclassical analysis
should be valid.  The superpotential cannot
be generated in perturbation theory.
In perturbation theory, there is an
additional $U(1)$ symmetry which forbids
a superpotential all together -- this is the
content of the standard non-renormalization theorem.
Alternatively (and equivalently) we note that in perturbation
theory only logs of $\Lambda$ appear, not powers.
In weakly coupled theories,
the only non-perturbative effects which we understand
are instantons.  Without actually doing a computation,
one can see that the instanton amplitude potentially has
the correct form.  Instanton effects go as
\beq
e^{-{8 \pi^2 \over g^2(v)}}
\label{instantonamplitude}
\eeq
where $g^2(v)$ is the coupling at the scale $v$.
In the present case, noting that the leading term in the
$\beta$ function goes as $2N+1$, 
\beq
e^{-{8 \pi^2 \over g^2(v)}}= (\Lambda/v)^{2N-1}
\label{running}
\eeq
which gives precisely the expected dependence on $\Lambda$.
I won't go through the details of the instanton computation
here.  They are reasonably straightforward \cite{instanton}.  Suffice
it to say that in the end, one obtains a contribution to the
effective action of the expected form, with a non-zero
coefficient.

For $N_f < N_c-1$, the gauge group is not completely broken,
even in the most general flat direction.  Instead, one is left
with the mesons, $M$, and an unbroken gauge group,
$SU(N-N_f)$, with no matter fields (i.e., a ``pure supersymmetric
gauge theory").  The gauge theory by itself presumably
confines and has a mass gap, so our goal, again, is
to obtain the effective theory of the light fields $M$.
The superpotential expected from eq.~(\ref{wnp})
is again generated, now as a result of gluino condensation
in the $SU(N-N_f)$ theory.  To understand how this works,
it is simplest to consider first the simplest case,
$SU(3)$ with a single flavor.  Then in the
general flat direction,
\beq
Q = \overline Q = \left ( \matrix{0 \cr 0 \cr v} \right ).
\label{suthreeflat}
\eeq
The low energy theory is an $SU(2)$ gauge
theory with an additional, neutral singlet field,
$M= \overline Q Q$.  The effective action can be organized into
terms with higher and higher powers of $1/v$.  In leading order,
one has a free, decoupled chiral multiplet and a pure $SU(2)$
gauge theory.  General arguments indicate that the
$SU(2)$ theory does not break supersymmetry, and
posesses a mass gap \cite{wittendsb,wittendsba}.

The lowest order term which couples the two sectors
involves the $f$ function of eq.~(\ref{nonrenormalizable}).
In the present
case, the form of this
function can be determined by symmetry considerations.
The microscopic theory possesses an $R$ symmetry,
under which
\beq
\lambda \rightarrow e^{i \alpha \lambda}\quad \quad
Q, \overline Q \rightarrow e^{-2 i \alpha} Q \overline Q.
\label{suthreesymm}
\eeq
The low energy theory contains only the gluinos
and this symmetry now seems to be anomalous.
The anomaly, however, will be cancelled if
the theory includes a coupling
\beq
{1 \over 16 \pi^2} \int d^2 \theta
\ln(\overline Q Q)W_{\alpha}^2
={1 \over 16 \pi^2}\int d^2 \theta \ln(M) W_{\alpha}^2.
\label{mwsquared}
\eeq
A simple one loop calculation verifies that this coupling
is indeed present.

In the pure gauge theory, there is a non-zero
gaugino condensate,
\beq
\langle \lambda \lambda \rangle
= \Lambda_{SU(2)}^3.
\label{gauginocondensate}
\eeq
Here, $\Lambda_{SU(2)}$ is the scale of the low energy
$SU(2)$ theory; it is related to the original $SU(3)$ scale
by 
\beq
\Lambda_{\rm su(2)}^3= v^3 e^{-{8 \pi^2 \over b_o^{\prime} g^2(v)}},
\label{matching}
\eeq
with $b_o^{\prime}=6$, the low energy $\beta$ function,
and
\beq
{8 \pi^2 \over  g^2(v)} = {8 \pi^2 \over
g^2(M)}+b_o \ln(v/M),
\label{gofv}
\eeq
where $b_o=5$ is the microscopic $\beta$ function.

Examining eq.~(\ref{mwsquared}), it is clear that
the gluino condensate gives rise to a superpotential
for $M$.  It is not hard to check that this has precisely
the correct form (it is easiest to do this in terms
of the component fields, because of the dependence of
the condensate on the fields).

So we have seen two possible behaviors for theories which
classically have moduli.  Either no potential is generated,
and the moduli are exact even quantum mechanically.
Alternatively, a potential is generated, but it falls to
zero in the region where the coupling constants tend to
zero.  These results are not suprising.   Later, we will
discuss an example in which a flat direction is lifted
and there is a stable ground state.  This occurs because
the coupling grows as the field becomes large.

We turn now to theories without classical flat
directions.  In such cases, the generic behavior is that
supersymmetry is unbroken.  This was already implicit
in our discussion of the pure $SU(N)$ gauge theory.
Consider, now, supersymmetric QCD with masses for the
quarks.  In particular, suppose that the mass is very
small.  Then there are still approximate flat directions,
and we expect that the superpotential is simply a sum
of the tree level (mass) term and the non-perturbative
term.  One can, in fact, prove that this form is exact,
using symmetries and holomorphy.  So, for example, for
$N_f=N-1$,
\beq
W= {\Lambda^{2N+1} \over \det \overline Q Q}
+ m \overline Q Q.
\label{wwithmass}
\eeq
Now there are no flat directions, classically or quantum
mechanically.  To make things simple, we have
taken all of the quark masses equal.
We can look for a supersymmetric minimum
by assuming that $Q$ has the form
\beq
Q = \left( \matrix{v_1 & 0 & \dots & 0 \cr
0& v_2 & \dots & \dots \cr \dots & \dots & \dots & \dots \cr
0 & 0 & \dots & v_{N_f} \cr 0 & 0 & \dots & 0 } \right).
\label{qminimum}
\eeq
Then the equation ${\partial W \over \partial Q}=0$
gives that
\beq
v^{2N} = {\Lambda^{2N+1} \over m},
\label{minwithmass}
\eeq
i.e., there are $N$ supersymmetry preserving roots.\footnote{This
is compatible with the computation of the Witten
index \cite{wittendsba}.}

\smallskip
\smallskip
\noindent
{\bf Exercise:}  Prove the existence of a gluino
condensate in pure $SU(N)$ gauge theory
by the following indirect argument \cite{seibergnr}.
Start with the case $N_f=N-1$, in which a direct
calculation leads to the superpotential expected from
eq.~(\ref{wnp}).  Add a mass term for one of the quarks, $Q_{N_f}$,
and integrate it out, by solving the equation
$${\partial W \over \partial Q_{N_f}}=0.$$
Substitute this solution back into the superpotential to
obtain an effective superpotential for the remaining
fields.  By holomorphy, this expression is guaranteed
to be correct for large $m_{N_f}$.  Verify that
the result is that expected from eq.~(\ref{wnp}).   Note,
in addition to the correct dependence on fields, this
also has the correct dependence on the scale $\Lambda$.
This demonstrates the existence of the gluino condensate,
since we have seen that such a condensate is required
to explain directly the existence of the non-perturbative
superpotential.

\smallskip
\smallskip

A model in which supersymmetry turns out to
be broken is the ``3--2 model."  This theory has gauge
symmetry $SU(3) \times SU(2)$, and matter content:
\beq
Q(3,2) \quad \quad \bar u(\bar 3,1) \quad
\quad L(1,2) \quad \quad
\bar d(\bar 3,1).  
\label{matterthreetwo}
\eeq
This is similar to the field content of a single generation
of the standard model, without the extra $U(1)$ and the
positron.
The most general renormalizable
superpotential consistent with the symmetries
is
\beq
W= \lambda Q L \bar u.
\label{threetwow}
\eeq
This model admits an R symmetry which is free of anomalies.
There is also a conventional $U(1)$ symmetry, under which
the charges of the various fields are the same as in the
standard model (one can gauge this symmetry if one
also adds an $e^+$ field).

While this model has global symmetries,
it is different from supersymmetric QCD
in that it does not
have classical flat directions.  To see this, note that by
$SU(3) \times SU(2)$ transformations, one can bring
$Q$
to the form
\beq
Q = \left ( \matrix{v_1 & 0 \cr 0 &  v_2 \cr 0  & 0 }
\right ).
\label{qvev}
\eeq
Suppose $v_1$ and/or $v_2$ is non-vanishing.  Then
if $v_1 \ne v_2$, vanishing of the $SU(2)$ $D$-term requires
that $L$ is non-zero.  This, however, implies that
$\partial W \over \partial \bar u$ is non-zero.
So $v_1$ must equal $v_2$, and $L$ must vanish.
But now one cannot make the $SU(3)$ $D$-term vanish
unless both $\bar u$ and $\bar d$ are non-vanishing,
in which case $\partial W \over \partial L$ is non-vanishing.
So there are no flat directions in this model.

To analyze the dynamics of this theory, consider first the
case that $\Lambda_3 \gg \Lambda_2.$  Ignoring,
at first, the superpotential term, this is just
$SU(3)$ with two flavors.  In the flat
direction of the $D$ terms,  there is a non-perturbative
superpotential,
\beq
W_{np}= {\Lambda^5 \over \det \overline Q Q}
\sim {1 \over v^4}.
\label{wnpthreetwoa}
\eeq
The full superpotential in the low energy theory is a sum
of this term and the perturbative term.  It is straightforward
to minimize the potential, and establish that
supersymmetry is broken.  The case that $\Lambda_2
\gg \Lambda_3$ has been analyzed more recently
by Intriligator and Thomas.  In this case, before including
the classical superpotential, the theory is $SU(2)$ with
two flavors.  This is an example of a model with
a ``quantum moduli space" (see Peskin's lectures at this
school).  Studying the effects of the classical superpotential
on this space, one again finds that supersymmetry is broken.

Even without detailed dynamical analysis,
we could have anticipated that supersymmetry
would be broken in this model
from the following argument.  This model has
no flat directions.  It also possesses a non-anomalous
global symmetry.  One expects that this global
symmetry is spontaneously broken.
In the limit of large $\Lambda_3$,
since the superpotential blows up at the
origin, the minimum
of the potential must lie away from the origin.  In the limit
of large $\Lambda_2$, the point of unbroken symmetry
does not lie in the quantum moduli
space.  This means that the $R$ symmetry
is spontaneously broken.  Correspondingly, there
must be a Goldstone boson.  If supersymmetry were
{\it unbroken} at the minimum, this field would necessarily
be part of a chiral multiplet.  The other scalar in this multiplet,
like the Goldstone particle, would have no potential, and
thus the full non-perturbative theory would have a flat
direction.  But this is essentially impossible, given that
all of the couplings are weak and the classical theory
has no flat directions.

This argument is often useful in considering models
which are inherently strongly coupled,
in which it is difficult to determine whether supersymmetry
is broken by direct calculation.  An example of this
kind is a model with gauge group $SU(5)$ and chiral fields
in the 10 and $\bar 5$ representations.  I will leave
it as an exercise to check that there are no flat directions,
and that there are two non-anomalous $U(1)$ symmetries.
While it is difficult to prove, one can at least advance
good arguments that one or both of these chiral symmetries
are broken, so again we expect that supersymmetry is
broken.  This model has an infinite set of generalizations:
theories with gauge group $SU(N)$, a single antisymmetric
tensor, and $(N-4) \overline N$'s.  All of these models are
believed to break supersymmetry.

\smallskip
\smallskip
\noindent
{\bf Exercise:}  Check that these models
have no flat directions, and that they have
a non-anomalous $U(1)$ symmetry.
 \smallskip
\smallskip

During the past year, many generalizations of these
models have been constructed \cite{dnns,newmodels}.
In addition,
new mechanisms of supersymmetry breaking
have been discovered \cite{intt,iss,nelsonnew}.
As an example,
consider a model with gauge group $SU(2)$
and four doublets, $Q_I,I=1 \dots 4$ (two ``flavors") \cite{intt}.
Classically,
this model has a moduli space labeled by the expectation
values of the fields
$M_{IJ} = Q_I Q_J$.   These satisfy
 ${\rm Pf}\langle M_{IJ} \rangle =0$, but,
as we noted in our discussion of the 3--2 model when
the same structure arose,
the quantum moduli space is different, and satisfies:
\beq
\rm Pf \langle M_{IJ} \rangle = \Lambda^4.
\label{quantummoduli}
\eeq

Now add a set of singlets to the model, $S_{IJ}$, with
superpotential couplings
\beq
W= \lambda_{IJ} S_{IJ} Q_IQ_J.
\label{itw}
\eeq
Unbroken supersymmetry now requires
\beq
{\partial W \over \partial S_{IJ}}= Q_IQ_J=0.
\label{itsusy}
\eeq
However, this is incompatible with the quantum constraint.
So it would appear that supersymmetry is broken.

On the other hand, the model, classically, has flat directions
in which $S_{IJ}=s_{IJ}$, and all of the other
fields vanish.  So one might worry that there is runaway
behavior in these directions, similar to that we saw
in supersymmetric QCD.  However, for large $s$,
it turns out that the energy grows at infinity \cite{yuri}.
This can be established as follows.  Suppose all
of the components of $S$ are large, $S \sim s \gg \Lambda_2$.
In this limit, the low energy theory is a pure $SU(2)$
gauge theory.  In this theory, gluinos condense,
\beq
\langle \lambda \lambda \rangle
= \Lambda_{LE}^3 = \lambda s \Lambda_2^2.
\label{gluinoit}
\eeq
Here, $\Lambda_{LE}$ is the $\Lambda$ parameter
of the low energy theory.

At this level, then, the superpotential of the model
behaves as
\beq
W_{\it eff} \sim \lambda S \Lambda_w^2,
\label{wit}
\eeq
and the potential is a constant,
\beq
V = \vert \lambda_2 \vert^4 \vert \lambda \vert^2.
\label{itv}
\eeq
However, with a little thought, it is clear that one
should think of $\lambda$ in this expression as the effective
$\lambda$ at the scale $s$.The behavior of $\lambda$
with $s$ depends on the values of the couplings, but
for a range of parameters $\lambda$ grows with $s$.
In some cases, it has a minimum at large $s$, where
the theory is weakly coupled.  Then one can determine
the location of the minimum and the pattern of symmetry
breakdown.  
In other cases, the minimum occurs in the region of small
$s$, where the theory is strongly coupled and difficult
to analyze.

We have seen, in this section, that dynamical breaking
of supersymmetry is common.  Flat directions are
often lifted, and in many instances,
supersymmetry is broken with a stable ground state.
So we are ready to address the question:  how might
supersymmetry be broken in the real world?

\section{Where is the Scale of Supersymmetry Breaking?}

If supersymmetry has something to do with the
hierarchy problem, it must, in some sense, be
broken at a scale of order $M_Z$.  More precisely,
the soft breakings among the ordinary quarks,
leptons and gauge particles must be of this order.
However, the fundamental scale of supersymmetry
breaking -- the scale  of the $F$ or $D$
fields which break supersymmetry -- can
be much larger.  Virtually all existing models
of supersymmetry breaking -- whether based on
dynamical breaking, or on tree level breaking
as in the O'Raifeartaigh model -- assume that some
new set of fields and interactions are responsible for symmetry
breakdown.  The most popular approach
has been to assume that supersymmetry is broken
in a ``hidden sector," i.e., by fields which have only
very tiny couplings to ordinary matter, and that the
breaking of supersymmetry is fed down to ordinary
fields by gravitational strength interactions.  In this
case, the scale of breaking is intermediate between the
Planck scale and the weak scale, of order
$10^{11}$ GeV.  Such an approach has a certain
degree of elegance, and is suggested by string theory.
As we will see, however, it is not easy to understand
how the problems raised by the rare processes we have
discussed earlier are resolved in this framework.
An alternative possibility is that the breaking
occurs at much lower scales, and is mediated by
gauge interactions.  This approach is remarkably
predictive, and automatically avoids the problems
of flavor changing neutral currents.  In this section,
we will review both of these possibilities.

\subsection{Hidden Sector N=1 Supergravity}

It would take many lectures (and a more expert
lecturer) to give a proper exposition of $N=1$ supergravity.
Fortunately, there are only a few facts we will need
to know.   First, the terms in the effective
action with at most two derivatives
or four fermions are completely specified by
three functions:
\begin{enumerate}
\item  The Kahler potential, $K(\phi,\phi^{\dagger})$,
a function of the chiral fields
\item  The superpotential, $W(\phi)$, a holomorphic
function of the chiral fields.
\item  The gauge coupling functions,
$f^a(\phi)$, which are also holomorphic functions of the
chiral fields.
\end{enumerate}

The lagrangian which follows from these can be found,
for example, in \cite{cremmer,nilles}.
Let us focus, first, on the scalar potential.  This is given
by
\beq
V=e^{K}\left[
\left({\partial W \over 
\partial \phi_i} + {\partial K \over \partial \phi_i} W
\right)
g^{i \bar j}
\left({\partial W^* \over 
\partial \phi^*_j} + {\partial K \over \partial \phi^*_j}W
\right )
- 3 \vert W \vert^2\right],
\label{supergravity}
\eeq
where
\beq
g_{i \bar j} = {\partial^2 K \over \partial \phi_i
\partial \phi_{\bar j}}
\label{kahlermetric}
\eeq
is the Kahler metric associated with the Kahler potential.
In this equation, we have adopted
units in which
\beq
G_N={1 \over 8 \pi  M^2}.
\label{reducedmp}
\eeq
where $M \approx 2 \times 10^{18}$ GeV is
the reduced Planck mass we encountered earlier.

There is a standard strategy for building
supergravity models.  One introduces two sets of fields,
the ``hidden sector fields," which will be denoted by $Z_i$,
and the ``visible sector fields," denoted $y_a$.  The $Z_i$'s
are assumed to be connected with supersymmetry breaking,
and to have only very small couplings to the ordinary fields,
$y_a$.  In other words, one assumes that the superpotential,
$W$, has the form
\beq
W= W_z(Z) + W_y(y),
\label{whideen}
\eeq
at least up to terms suppressed by $1/M$.

One also usually assumes that the Kahler potential
has a ``minimal" form,
\beq
K=\sum z_i^{\dagger} z_i +
\sum y_a^{\dagger} y_a.
\label{minimalkahler}
\eeq
One chooses (tunes) the parameters of $W_Z$ so that
\beq
\langle F_Z \rangle \approx M_w M
\label{fzis}
\eeq
and
\beq
\langle V \rangle = 0.
\label{vanishingv}
\eeq
Note that this means that
\beq
\langle W \rangle \approx M_W M_p^2.
\label{meanw}
\eeq

To understand the structure of the low energy theory in such
a model, suppose first that $W(y)=0$, and that
$K$ is of the ``minimal" form, eq.~(\ref{minimalkahler}).  Then
\beq
V(y) = e^{K(z)}\vert \langle
W \rangle \vert^2   \sum \vert y_a \vert^2.
\label{vofy}
\eeq
In other words, the scalars all have a common
mass,
\beq
m_o^2= e^{\langle K \rangle}
{\vert \langle W \rangle \vert^2 \over M^4}
\approx m_{3/2}^2.
\label{scalarmassint}
\eeq
Here $m_{3/2}$ is the mass of the gravitino.
Note that with $F \sim  M
M_Z$, $m_{3/2} \sim M_Z$.

If we now allow for a non-trivial $W_y$, we find also
$A$ and $B \mu$ terms.  For example,
the terms
\beq
{\partial W \over \partial y_a} y_a W_y = 3 W_y
\label{partialyy}
\eeq
if $W$ is homogeneous of degree three.  Additional
contributions arise from
\beq
\left\langle {\partial W \over \partial z_i} \right\rangle
y_j^* W^* + \rm c.c.
\label{partialzy}
\eeq

\smallskip
\smallskip
\noindent
{\bf Exercise}:  The simplest model of the hidden
sector is known as the ``Polonyi model." In this model,
\beq
W=m^2(z+ \beta)
\label{wpolonyi}
\eeq
\beq
\beta = (2+ \sqrt{3} M) \ .
\label{betais}
\eeq
Verify that the minium of the potential
for $Z$ lies at
\beq
Z= (\sqrt{3}-1) M
\label{zpolonyi}
\eeq
and that
\beq
m_{3/2} = (m^2 /m) 
e^{(\sqrt{3}-1)^2/2} \quad  m_o^2=2 \sqrt{3} m_{3/2}^2
 \quad  A=(3-\sqrt{3}) m_{3/2}.
\label{poloniparams}
\eeq
\smallskip
\smallskip

So far, we have not addressed the question of gaugino
mass.  This can arise from a non-trivial gauge coupling
function,
\beq
f^a = c{Z\over M}
\label{gauginois}
\eeq
which gives a gluino mass, just as it would in the
global case:
\beq
m_{\lambda}= {c F_z \over M}.
\label{gluinomassis}
\eeq

So these models have just the correct structure.  They have
soft breakings of the correct order of magnitude,
and they exhibit, with our assumption of minimal
kinetic term, the properties of universality and
proportionality.  Indeed, this
is a highly predictive framework, with only (assuming
MSSM particle content) $5$
parameters.  A large amount of work has
been done on these models, including investigations of:\footnote{More
detail on all of these points is provided in many excellent
review articles.  See, for example, Jon Bagger's 1995 TASI
lectures, and references therein \cite{bagger}.}
\begin{enumerate}
\item  Renormalization group evolution:  one finds
that there is a substantial region of the parameter
space in which $SU(2)\times U(1)$ is broken
in the correct fashion.
\item Unification:  The predictions for unification
are far better than in the minimal standard model.
However, the prediction for $\alpha_s$ tends to
be somewhat larger than observed.  This can be ``fixed"
by adding new thresholds at the high
scale \cite{damien}.
\item  $b$-$\tau$ unification: One can also consider
the possibility that the $b$ and $\tau$ Yukawa's unify.
This is certainly not a general requirement of unification;
it depends, in grand unified models, on the Higgs content,
and in string theory on the details of compactification.
Still, this idea seems viable (and interesting).
\item  Proton decay:  with detailed assumptions
about the susy spectrum and the structure of grand
unification, it is possible to compute the proton lifetime and
compare with experimental limits.
\item  Dark matter:  Again, in a detailed model,
one can identify the LSP and compute its couplings to
matter.  This permits a precise calculation of the abundance.
\item  $b \rightarrow s \gamma$:  As we have already
remarked, one can compute the rate for this process in particular
models.  One often finds that other constraints are stronger.
\end{enumerate}

\subsection{Beyond the Minimal Model}

Almost all work on supersymmetry phenomenology
has been based on the idea that supergravity
is the messenger of supersymmetry breaking, and
the assumption that, at the high scale, the theory
is described by an effective lagrangian with a
Kahler potential of the form of eq.~(\ref{minimalkahler}).
Yet this assumption is hard to justify.  It is often said
that it is plausible, since gravity is ``flavor blind."
But the flavor-blindness of gravity is a consequence
of general covariance; no such principle forbids a Kahler
potential of the form:
\beq
K=f(Z,Z^{\dagger})+ \sum y_a^{\dagger}
y_a + \sum_{ab} h_{ab}(Z,Z^{\dagger})\phi_a^{\dagger}\phi_b + \dots\ .
\label{knonuniversal}
\eeq
This last term is similar to terms we described
in the context of global supersymmetry, and has
{\it exactly} the same effect:  it spoils universality and
proportionality.  These terms are not forbidden by
any symmetry, and they are generated by
radiative corrections, so they are
surely present at some level.

$N=1$ supergravity models are not renormalizable,
and thus can't be viewed as in
any sense complete, predictive theories.
Without some understanding of the underlying
microscopic theory, there is little more one can say.
One possibility is that the underlying theory possesses
some flavor symmetry.  This is presumably a gauge
symmetry (continuous or discrete), since it is widely
believed that global symmetries do not make sense
in the framework of quantum gravity.  These symmetries
must be broken, and it is a challenge to simultaneously
obtain sufficient squark and slepton degeneracy
while accomadating the strong violation of flavor
observed in the fermion sector \cite{flavorsymmetries}.

Ultimately, however, the only candidate we have for
a microscopic theory of supergravity is string theory.
In principle, it should be possible to calculate the Kahler
potential in string theory and determine whether it yields sufficient
degeneracy.  Some tentative steps in this direction
have been taken, and the results are mixed.  There are
limits in which string theory does yield the required
structure.  On the other hand, it is hard to see how
these limits could have anything to do with the real
world.  We will return to this question later, in the
chapter on string theory.

\subsection{Incorporating Dynamical Supersymmetry Breaking}

Models like the Polonyi model are extremely ad hoc.  They
introduce a small parameter by hand, subject to
strong constraints.  It would be much more satisfying
if one had a dynamical understanding of this new scale.
We have seen that dynamical supersymmetry breaking (DSB)
occurs in many theories, and in such theories the
appearance of a small scale is easily explained as an
effect of order $e^{-c/g^2}$, for some weak coupling $g$.

There are two approaches which have been
tried to incorporating DSB into the framework
of supergravity models.   The first is to replace
the Polonyi sector with a model, such as the 3--2
model, in which supersymmetry is broken and there
is a stable ground state.  
Attempts to do
this, however, run into difficulty producing a phenomenologically
acceptable gluino mass.  The problem is that
in the simplest theories, there are no gauge singlet
superfields in this hidden sector, so gluino masses
can only arise through operators of high dimension,
such as
\beq
\int d^2 \theta {\Phi^3 \over M_p^3}W_{\alpha}^2.
\label{highdgluino}
\eeq
If the typical scale of the hidden sector is of order $10^{11} \rm
GeV$, this generates a gluino mass of order $10^{-7}$ eV
or so (actually, larger contributions will be generated by loops
in the very low energy, non-supersymmetric theory, but these
will not be nearly large enough).  

It is possible to write DSB models with gauge singlets; the model
of Intriligator and Thomas which we discussed earlier is an
example.  In the construction of these models, however, it is crucial
that there are no terms of order $S^2$ or $S^3$.  One might
try to explain the absence of such terms by discrete symmetries.
But such symmetries inevitably forbid the desired linear
term in $S$ in the gauge coupling function.
Theses models also offer no further insight into the question
of universality.  No symmetry will forbid the dangerous
terms in the Kahler potential of eq.~(\ref{knonuniversal}). 

The second framework in which to consider this question
is string theory.   Since all known
string models possess moduli, it is necessary to understand
how the degeneracy among the associated vacuum states
is lifted.  As we will discuss in the chapter on string
theory, there are always directions in which the potential
for the moduli tends to zero.  In fact, it is not hard to
argue that the potential for the moduli can only have
local minima in regions where a perturbative or semiclassical
analysis breaks down.  If one assumes the moduli are stabilized,
then the gaugino and scalar masses are comparable.
However, one still has difficulty understanding
degeneracy and proportionality.  We will discuss these issues
in some detail in section 7.

\section{Low Energy Dynamical Supersymmetry Breaking}

\begin{figure}[htbp]
\centering
\centerline{\psfig{file=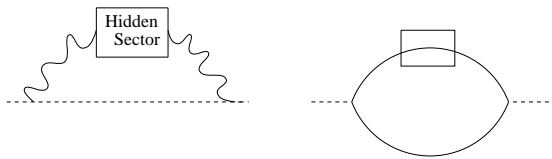,height=2cm,angle=-90}}
\caption{Schematic of gauge mediated supersymmetry
breaking.
}
\label{basicgmsb}
\end{figure}

An alternative to the conventional supergravity
approach is to suppose that supersymmetry is broken
at some much lower energy, with gauge interactions
serving as the messengers of supersymmetry
breaking \cite{dn,dn2,dnns}. The basic
idea is indicated in fig.~\ref{basicgmsb}.  One again supposes that
one has some set of new fields and interactions which break
supersymmetry.  Some of these fields are taken
to carry ordinary
standard model quantum numbers, so that
 ``ordinary" squarks, sleptons and gauginos
can couple to them through gauge loops.  This
approach, which is referred to as ``gauge mediated
supersymmetry breaking" (GMSB) has a number of virtues:
\begin{enumerate}
\item  It is highly predictive:  as few as $2$ parameters
describe all soft breakings.
\item  The degeneracies required to suppress flavor
changing neutral currents are automatic.
\item  GMSB easily incorporates DSB, and so
can readily explain the hierarchy.
\item  GMSB makes dramatic and distinctive
experimental predictions.
\end{enumerate}

The approach, however, also has drawbacks.
Perhaps most serious is related to the ``$\mu$ problem,"
the
question:  why is the $\mu$-parameter of the MSSM
of order $M_Z$ rather than, say $M_p$ or any
scale in between.  
I have not alluded to this problem earlier, because
in the supergravity framework it is not a really a problem
at all.\footnote{I should note that this statement is
not completely non-controversial.}In string theory,
one can give the following answer.
It is quite common to find states which are massless
at tree level, even though there is no symmetry
which explains the absence of a mass term.  In a
non-supersymmetric theory, the mass would be corrected
in loops, but now the non-renormalization theorems
forbid a mass to any order.  So the fact that $\mu$
is very small is not suprising.  Assuming
that supersymmetry is broken at an intermediate
scale in the desired fashion, supersymmetry breaking
effects tend to generate a $\mu$ term of precisely
the correct order, through couplings like
\beq
\int d^4 \theta {Z^{\dagger}\over
M}H_UH_D.
\label{muterm}
\eeq
This phenomenon actually occurs in the simplest
supergravity models \cite{arnowitt,agpw}.

The $\mu$ problem, however, finds a home in the
framework of low energy breaking.
The difficulty is that, if one is trying to
explain the weak scale dynamically, one does
not want to introduce the $\mu$ term by hand.
However, operators like those of eq.~(\ref{muterm})
now do not generate a $\mu$ term of the correct
order of magnitude.  Various solutions have been offered
for this problem \cite{dnns,pomarol},
but none is yet compelling.
Perhaps the most likely possibility is that there will
be some structure beyond that of the MSSM at relatively
low energies, such as singlets coupled
to Higgs fields.  In most of our discussion, we will
simply assume that a $\mu$ term has been generated
in the effective theory, and not worry about its origin.

\subsection{Minimal Gauge Mediation (MGM)}

\begin{figure}[htbp]
\centering
\centerline{\psfig{file=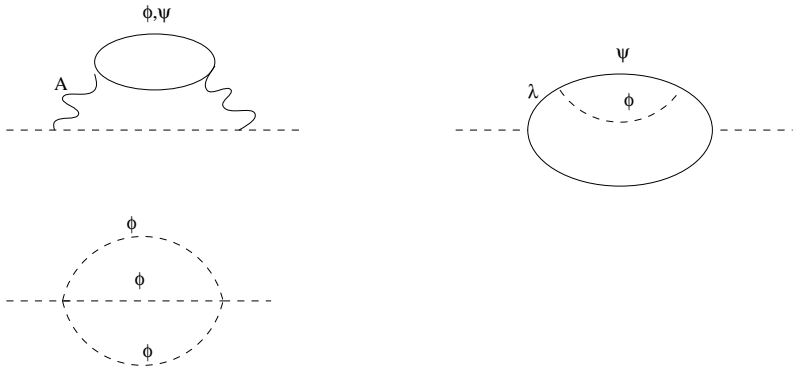,width=10.5cm,angle=-90}}
\caption{Two loop diagrams contributing to squark
masses in a simple model of gauge mediation.}
\label{twoloopgraphs}
\end{figure}

The simplest model of gauge mediation contains,
as messengers, a vectorlike set of quarks and leptons,
$q$, $\bar q$, $\ell$ and $\bar \ell$.  These have
the quantum numbers of a $5$ and $\bar 5$
of $SU(5)$.  The superpotential is taken to be
\beq
W_{mgm}=\lambda_1 q \bar q
+ \lambda_2 S \ell \bar \ell.
\label{mgmw}
\eeq
We suppose that some dynamics gives rise to non-zero
expectation values for $S$ and $F_S$.  We will
not, in these lectures, explore detailed proposals
for this dynamics; for this see \cite{dnns}.
Instead, we will go ahead and immediately compute
the superparticle spectrum for such a model.
Ordinary squarks and sleptons gain mass through
the two-loop diagrams shown in fig.~\ref{twoloopgraphs}.  While
the prospect of computing a set of two
loop diagrams may seem intimidating, the computation
is actually quite easy.  If one treats $F_S/S$
as small, there is only one scale in the integrals.
It is a straightforward matter to write down the
diagrams, introduce Feynman parameters, and
perform the calculation.  There are also various
non-trivial checks.  For example, the sum of the
diagrams must vanish in the supersymmetric limit.

One obtains the following expressions for
the scalar masses \cite{dn2,dnns}:
\beq
\widetilde m^2 ={2 \Lambda^2} 
\left[
C_3\left({\alpha_3 \over 4 \pi}\right)^2
+C_2\left({\alpha_2\over 4 \pi}\right)^2
+{5 \over 3}{\left(Y\over2\right)^2}
\left({\alpha_1\over 4 \pi}\right)^2\right],
\label{scalarsmgm}
\eeq
where  $\Lambda=F_S/S$,
and $C_3 = 4/3$ for color triplets and zero for singlets,
$C_2= 3/4$ for
weak doublets and zero for singlets.
For the gaugino masses one obtains:
\beq
m_{\lambda_i}=c_i\ {\alpha_i\over4\pi}\ \Lambda\ .
\label{gauginomass}
\eeq
This expression is valid only to lowest order in $\Lambda$.
Higher order corrections have been computed in \cite{higherorder}.

All of these masses are positive, and
they are described in terms of a single
new parameter, $\Lambda$.  The lightest
new particles are the partners of the
$SU(3)\times SU(2)$ singlet leptons.
If their masses are of order 100 GeV,
we have that $\Lambda \sim 30 ~{\rm TeV}$.
The spectrum has a high degree of
degeneracy.  In this
approximation, the masses of the squarks and sleptons
are functions only of their gauge quantum numbers,
so flavor changing processes are suppressed.
Flavor violation only arises through Yukawa couplings,
and these can only appear in graphs at high loop
order.  It is further suppressed because
all but the top Yukawa coupling is small.

Apart from the parameter $\Lambda$, one has
the $\mu$ and $B \mu$ parameters (both complex), for a total of
five.  This is three beyond the minimal standard model.
If the underlying susy-breaking theory conserves CP,
this can eliminate the phases, reducing the
number of parameters by $2$.

\subsection{$SU(2) \times U(1) Breaking$}

At lowest order, all of the squark and slepton
masses are positive.  The large top quark Yukawa
coupling leads to large corrections to $m^2_{H_U}$,
however, which drive $SU(2) \times U(1)$ breaking.
The calculation is just a repeat of one we have done
earlier.  We can evaluate the diagram of fig.~\ref{oneloopyukawas},
treating the mass of $\tilde t$ as independent
of momentum, provided we cut the integral off
at a scale of order $\Lambda$ (at this scale,
the calculation leading to eq.~(\ref{scalarsmgm})
breaks down, and the
propagator falls rapidly with momentum) and we have
\beq
m^2_{H_U}=(m^2_{H_U})_o
-{6 y_t^2 \over 16 \pi^2}
\ln (\Lambda^2/\widetilde m_t^2)(\widetilde m_t^2)_o.
\label{twoonebreaking}
\eeq

While the loop correction is nominally a three loop
effect, because the stop mass arises from gluon loops while
the Higgs mass arises at lowest order from $W$ loops,
we have
\beq
\left ({\widetilde m_t^2 \over m^2_{H_U}} \right )_o
= {16 \over 9}({\alpha_3 \over \alpha_2})^2 \sim 20
\label{ratioth}
\eeq
and the Higgs mass-squared is negative.  This calculation
was in fact done many years ago in \cite{agcw}.
In some sense, the situation
here is more striking than in the supergravity case.
There, the soft breakings were all parameters anyway,
so the fact that there are negative radiative corrections to
some, while suggestive, does not permit a ``prediction" of
weak interaction symmetry breaking.  In the present case,
this is a prediction of the theory.

\subsection{Light Gravitino Phenomenology}

There are other striking features of these models.
One of the more interesting is that the LSP is the
gravitino.  Its mass is
\beq
m_{3/2}= 2.5 \left ({F \over (100\ {\rm TeV})^2} \right ) \rm
eV.
\label{gravitinomass}
\eeq
The next to lightest supersymmetric particle, or NLSP, can
be a neutralino, or a charged right handed slepton.
The NLSP will decay to its superpartner plus a gravitino
in a time long compared to typical microscopic times, but
still quite short.  The lifetime can be determined
from low energy theorems, in a manner reminiscent
of the calculation of the pion lifetime.  Just as the
chiral currents are linear in the (nearly massless)pion field,
\beq
j^{\mu 5}=f_{\pi} \partial^{\mu} \pi \quad \partial_{\mu}
j^{\mu 5}=\partial^2 \pi \approx 0
\label{pcac}
\eeq
so the supersymmetry current is linear in the Goldstino, $G$:
\beq
j^{\mu}_{\alpha}= F \gamma^{\mu} G +
\sigma^{\mu \nu} \lambda
F_{\mu \nu} + \dots
\label{nonlinearsusy}
\eeq
$F$, here, is the goldstino decay constant.
From this, if one assumes that the LSP is mostly photino,
one can calculate the amplitude for
$\tilde \gamma \rightarrow G+\gamma$ in
much the same way one considers processes
in current algebra.  From
eq.~(\ref{nonlinearsusy}), one sees
that $\partial_{\mu} j^{\mu}_{\alpha}$
is an interpolating field for $G$, so:
\beq
\langle G \gamma \vert \tilde \gamma \rangle
= \smallfrac 1 F \langle \gamma \vert \partial_{\mu}
 j^{\mu}_{\alpha} \vert \tilde \gamma
\rangle.
\label{goldstinoelement}
\eeq
The matrix element can be evaluated by
examining the second term
in the current, eq.~(\ref{nonlinearsusy}), and noting
that $\slash \partial \lambda = m_{\lambda} \lambda$.

Given the matrix element, the calculation of the NLSP lifetime is straightforward, and yields
\beq
\Gamma(\tilde \gamma \rightarrow
G \gamma)={\cos^2 \theta_W m_{\tilde \gamma}^5 \over
16 \pi F^2}.
\label{lifetime}
\eeq
This yields a decay {\it length}:
$$
c \tau = 130 \left({100\ \rm GeV \over m_{\tilde \gamma}} \right)^5
$$
\beq
 \left ( {\sqrt F \over 100\ {\rm TeV}}\right )^4 \mu m.
\label{ctau}
\eeq
\begin{figure}[htbp]
\centering
\centerline{\psfig{file=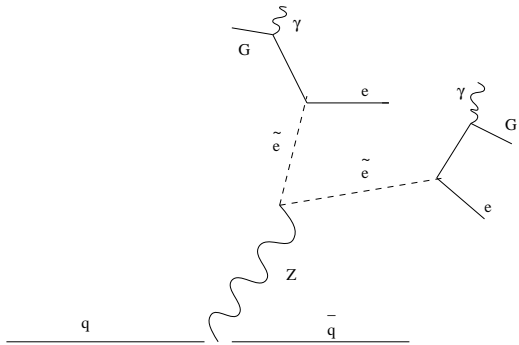,height=4.5cm,angle=-90}}
\caption{Decay leading to $e^+e^- \gamma \gamma$ events.
}
\label{eeggevents}
\end{figure}
In other words, if $F$ is not too large, the NLSP may
decay in the detector.  One even has the possibility
of measurable displaced vertices \cite{ddrt,stw}.
The signatures of such low decay constants would
be quite spectacular.  Assuming the photino (bino)
is the NLSP, one has processes such as $e^+e^-
\rightarrow \gamma \gamma + \slash  E_t$
and $p \bar b \rightarrow e^+e^- \gamma
\gamma +  \slash E_t$, as indicated in fig.~\ref{eeggevents}.
The first process has already
been used to set limits on the parameter space
of these models.
The second process may, conceivably, already
have been seen at the tevatron.  One event
has been seen with two electrons, two photons,
and substantial missing energy \cite{cdfdpf}.
Known standard model backgrounds for this
process are negligible.
One event is roughly consistent with the
expected cross section if the selectron mass is
in the 100~GeV range \cite{paigeetal}.
If the event is real, and this is the correct
explanation, one would expect to see many such events
at the upgraded tevatron.  The collider phenomenology
of these models has been studied in some
detail in~\cite{scottetal}.

It is natural to ask how general is the MGM.
Various modifications are possible, and these
have been discussed in \cite{ddrt,dns}.
Among the possible modifications are changes
in the particle content of the messenger sector,
and mixings of messenger and ``matter"
fields.

So far, while we have assumed
that supersymmetry is dynamically broken,
we have not examined the details of the
supersymmetry-breaking dynamics which
might underlie the MGM.
Indeed, perturbative models with the features
of MGM were constructed long ago, using ``O'Raifeartaigh"
type breaking \cite{dinefischler,nappiovrut,agcw}.
Realistic models in which supersymmetry
is dynamically broken {\it have} been
constructed \cite{dn,dn2,dnns}.  All of the constructions
so far involve a supersymmetry breaking sector
separated from the fields of the MSSM.
One might imagine that
we could do better, using the many new things we have
learned over the past three years.  One could
speculate on a theory with a very compact, tight
structure, in which, perhaps, some of the
gauge bosons we view as fundamental
are in fact composite, and the scales of 
supersymmetry and weak interaction symmetry
breaking are intimately tied
together \cite{seibergduality}.  Ref.~\cite{dns}\ makes
arguments that many features of this ultimate
theory are likely to be like those of the MGM.
Still, one worries that we are not being clever
enough.  One possible alternative viewpoint, which
starts by insisting on quite stringent -- and plausible --
notions of naturalness has been developed in \cite{ckn}.

\section{String Phenomenology and Phenomenology
of Strings}

Supersymmetry seems to be an essential
part of string theory.  Vast numbers of string
vacua are known with varying amounts of supersymmetry.
Much -- it is probably fair to say all -- of the
recent progress in string duality relies on
supersymmetry.
In fact, it is quite hard to make sense of string
vacua which don't possess at least some degree
of supersymmetry.  In particular, there seems
invarariably to be a cosmological constant
at one loop.  In addition to the fact that ``constant"
is far too large, what one is really determining
in these computations is a potential for
the dilaton.  This potential always tends
to zero for very weak coupling, which means that the
vacuum is unstable \cite{cosmostring}.

The methods of the previous sections can
be applied to understand many issues in string
phenomenology.  But they can also be used
to give insights into the basic dynamics of string theory.
In this section, we will content ourselves with
a brief survey of these ideas.

At the classical level, 
moduli are ubiquitous in string
theory.  They are problematic for any string
phenomenology.  One wants to understand
how a particular vacuum state is selected,
and how the moduli gain mass.   At the same time, they
provide tools to understanding many aspects of string
dynamics.  Indeed, this is a large part
of the duality story.

A simple example of this bad/good aspect of moduli
is provided by toroidal compactification of the
heterotic string to four
dimensions \cite{narain}.
These possess $N=4$ supersymmetry.  There
is a large moduli space.   At weak
coupling and low energies, one can write
an effective action for the light degrees of freedom --
the moduli, the gauge fields and the graviton
supermultiplet.  This effective action
must respect the full supersymmetry.
But $N=4$ supersymmetry is so restrictive that it forbids any
terms in ${\cal L}_{\it eff}$ which might lift the
degeneracy.  This statement holds perturbatively
and non perturbatively!  In other words,
the moduli in these theories are exact.
There are also lots of $5,6\dots 10$ dimensional
vacua which are exact.  To date, we have no
idea what might provide a vacuum selection principle,
which would explain why we live in an approximately
$N=1$ supersymmetric vacuum.
Indeed, it is not easy to understand why there
should even exist a vacuum with $N=1$ supersymmetry
broken to $N=0$, never mind explaining how the
universe finds itself in such a state.

We will focus on a particular class of potential string vacua:
perturbative ground states of the heterotic string
theory with $N=1$ supersymmetry.  All known
states of this type possess moduli.  One of these
controls the size of the four dimensional gauge
couplings at the classical level; it is known as the
``model-independent dilaton." It is convenient to
write the (complex) scalar component of this multiplet
as
\beq
S= {8 \pi^2 \over g^2}+ia.
\label{dilatondefined}
\eeq
$S \rightarrow \infty$ corresponds to
weak string coupling if the other
moduli are held fixed.  Non-perturbative effects inevitably
fall to zero at weak coupling.
Analyticity of the superpotential and the
gauge coupling as a function of $S$, along
with certain symmetry properties, allow
us to make a number of strong statements
about the non-perturbative structure of the
theory.

First, we can prove a non-renormalization theorem
for perturbation theory similar to that we encountered
in field theory, and in almost the same way.
In perturbation theory, string theory has
a symmetry under the shift
\begin{equation}
a \rightarrow a+i \delta.
\end{equation}
This symmetry can be understood by examining string
vertex operators, or simply by noting that the
axion is related by duality to the antisymmetric tensor
field,
\beq
\partial_{\mu} a = \epsilon_{\mu \nu \rho\sigma}
H^{\nu \rho \sigma}.
\label{axionduality}
\eeq
In perturbation theory, the effective action can be written in terms
of $H$, as a consequence of the antisymmetric tensor
gauge invariance, so only derivates of $a$ appear.
Non-perturbatively, this Peccei-Quinn symmetry
is broken.  This is familiar from field theory,
where instantons break such symmetries.
However, as in field theory, a discrete subgroup
of the full group survives.  This is a consequence of
a quantization condition for the antisymmetric
tensor field \cite{rohmwitten}.
The coupling of a string to $B_{\mu \nu}$
to a string has the form:
\beq
I_B={1 \over 2} \int d^2 \sigma B_{MN}(\partial X^M
\partial X^N - M \leftrightarrow N).
\label{bstringaction}
\eeq
In order that string amplitudes be single-valued,
this action must be a multiple of $2 \pi n$.  If
$\Sigma$ is a closed three surface, this leads to
\beq
\int_{\Sigma} d^3 ~\Sigma H=n.
\label{bquantization}
\eeq
The Peccei-Quinn charge is $j_{\mu}^{PQ} = \partial_{\mu}a$,
so taking the divergence, and using the duality relation
between $a$ and $H$ leads to a condition for the change of the
Peccei-Quinn charge in any process \cite{coping}:
\beq
\Delta Q^{PQ}= \int d^4 x \partial^2 a = n.
\label{selectionrule}
\eeq
This means that, in the effective action, only operators
like $e^{ina}$ can appear.  So the
effective action is invariant under $2 \pi$ shifts of $a$.

This periodicity is a subgroup of ``S-duality."  Clearly, however,
it holds whether or not the particular string vacuum is S-dual.
Since the superpotential is analytic in $S$, and must respect
this periodicity property, it must be of the form:
\beq
W = w_o(M) + w_1(M)e^{-S} + w_2(M)e^{-2S} + \dots
\label{wcorrections}
\eeq
at least if the coupling is small.  As a consequence, $V$ goes rapidly to
zero as $S \rightarrow \infty.$  {\it Inevitably}, any
good vacuum of string theory lies at strong
coupling \cite{strongcoupling}.

A little thought indicates that the true vacuum is unlikely
to reside in any region where duality is very useful.
The problem is that if the strong coupling region is mapped
into some dual, weakly coupled theory, then
one will still have the problem of vacuum instability.
Unfortunately, we seem to be stuck with what David
Gross has dubbed ``The Principle of Minimal Calculability."
In other words, the true vacuum of string theory must
lie in a regime where no weak coupling analysis is applicable.
One can imagine loopholes in this argument, so I offer the
following exercise:

\smallskip
\smallskip
\noindent
{\bf Exercise}:  Find the loophole in the above
argument, find string vacua with
broken supersymmetry and vanishing cosmological
constant, and make first class reservations to Stockholm.
\smallskip
\smallskip

Actually, the situation is perhaps not so hopeless; holomorphy
of $W$ and $f$, and axion periodicity, might still permit us to make
definite predictions about the low energy theory.
Let us suppose that $S$ is as large as typical computations
of the unified coupling suggest,
\beq
\langle S \rangle ={8 \pi^2 \over g^2} \sim 150.
\label{svalue}
\eeq
We will give some arguments for this later.
In that case, $e^{-S}$ is an extremely small
number, and can be ignored.  In other words, the
superpotential is unchanged from its tree level
form, and the gauge couplings receive no
appreciable corrections beyond one loop.
However, the Kahler potential is not restricted
by holomorphy, and thus one expects it to receive
significant corrections at strong coupling.

One can conceive of corrections to $K$ such
that, even though the superpotential is of the form
of eq.~(\ref{wcorrections}), the potential has a local minimum
at $\langle S \rangle$.  From the perspective
of field theory, one might have expected perturbation
theory to be good for such small values of the
coupling.  But there are two arguments against this.
First, string
perturbation theory is not as convergent as field
theory perturbation theory, so perhaps the expansion of
the Kahler potential has already broken down for
such couplings.  Second, $S$ is not really the string
expansion parameter in any case.  The actual expansion
parameter is the expectation value of the $10$-dimensional
dilaton, and as we will discuss shortly, this can
be quite large even if the effective four dimensional
coupling is small.  In the following discussion, we will
assume that $e^{-S}$ is very small, but that corrections
to the Kahler potential are large, and are responsible
for stabilizing the potential for the dilaton and other
moduli.  Within our present understanding of string
theory, such a picture seems almost inevitable, if
the theory is to have anything to do with nature.

This assumption, plus the periodicity we discussed
earlier, has significant implications for the effective
action.  The effective action we are considering here
is the Wilsonian effective action, at a scale just below
the string scale.  In this action
\beq
W = W_{tree} + {\cal O}(e^{-S})
~~~~~f = f_{tree} + f_{1-loop} + {\cal O}(e^{-S})\ .
\label{smallcorr}
\eeq
This means that 
\begin{itemize}
\item  The spectrum of light states is the same at weak coupling
as at strong coupling.
\item  Yukawa terms in $W$ are the same as at weak
coupling, and so can be calculated at weak coupling.
Note that these are not the same as the physical Yukawa
couplings, which include terms in the Kahler potential,
which are renormalized.  In specific models, ratios of Yukawa
couplings, for example, may not be renormalized.
\item  The gauge couplings are unified.  Assuming that
the low energy theory contains only three generations,
without extra vector-like matter, then one obtains,
as we have discussed, reasonable agreement with experiment.
Note, however, that the unification scale computed in this
way is not necessarily associated with any physical threshold.
As one passes to strong coupling, there can be large
``renormalizations" of the masses of the massive states.
(In the framework of weakly coupled field
theories, the issues
described here have been considered in \cite{sv,shifman}.)
\end{itemize}

In a picture like this, anything having to do with the
Kahler potential cannot
be calculated.  This means that the soft breakings,
which we have seen depend crucially on the Kahler
potential, are not calculable unless one actually
solves the strong coupling problem.  This may seem quite
disappointing,
but with what we currently understand about
string theory, this is probably the best 
we can do.  However, I claim that if we were successful
(e.g., found a model and made a few successful quantitative
predictions) this would be an incredible triumph for
string theory.

Despite these negative comments about computing
the Kahler potential, I would like to mention a weak-coupling
scenario which, if it were somehow realized, would
be quite predictive and would give the sort of
universality usually assumed for supergravity
theories \cite{ibanezlust,vadim}.  In this
picture, it is assumed that the dilaton dominates
supersymmetry breaking, i.e., that $\langle F_S \rangle$
is by far the largest non-vanishing $F$-component.
It is also assumed that the weak coupling approximation
is valid.  In this case, the Kahler potential for $S$ is
known\cite{wittendr}
\beq
K=-\ln(S+S^{\dagger}) + K({\cal M},{\cal M}^{\dagger}).
\eeq
The assumption that the dilaton dominates supersymmetry
breaking means that $W=W(S)$.
There are no terms in $K$ at this order which couple
matter fields to $S$.  As a result, the squark
and slepton masses are universal.
It is straightforward
to calculate the soft breaking terms.
One has, for the scalar masses, gaugino masses, and
$A$ terms:
\beq
m_o^2= m_{3/2}^2~~~~~M= {\sqrt{3} \over 2}m_{3/2}
~~~~~A= -\sqrt{3}m_{3/2}.
\eeq

So, while we have argued that there is no reason to
think that a supergravity theory should yield universality
and proportionality, we see that there is a limiting case
of string theory which yields just that!  On the other
hand, in the limit that the coupling is small enough that
one can do perturbation theory, so that the expression
above is a good approximation to the Kahler potential,
we have argued one cannot expect to find a stable ground state
of string theory.\footnote{A possible loophole is
provided by racetrack models \cite{racetrack}.
However, efforts to construct such models
have not been particularly successful, either
in generating models with reasonable supersymmetry
breaking or in generating a phenomenologically
acceptable value of~$S$.}  One has to hope that,
accidentally, some Kahler corrections are large
and some are small.  Even if one assumes a good
vacuum, degeneracy may not hold to the required level
of accuracy \cite{louisnir}.
This model makes strong predictions for
the soft susy parameters, and it is not clear that these
are compatible with experimental constraints, at least
if the low energy theory has MSSM particle content \cite{louisdilaton}.

Returning to our picture of Kahler stabilization of the
dilaton, we have seen that at the high scale,
supersymmetry breaking effects are of order $e^{-S}$,
i.e., far too small to be of interest.  Effects in the
low energy theory, however, can be much
larger.  As an example, consider a Calabi-Yau
theory with the standard embedding of he gauge
group, and perhaps some Wilson lines.  Suppose that
there is a hidden sector with some gauge group
and associated one-loop $\beta$-function $b_o$.
Gluinos condense in this theory:
\beq
\langle \lambda \lambda \rangle = e^{(-3{8 \pi^2 \over b_o
g^2} -3i{ a \over b_o})}.
\label{lambdacondensate}
\eeq
This leads to an effective superpotential for $S$,
\beq
W_{\it eff}(S)= e^{-{3S \over b_o}}.
\label{weffs}
\eeq
It is possible to invent Kahler potentials which
lead to a minimum for the potential at $S_o$
with vanishing cosmological constant, and we will
assume that this is what occurs.
For suitable $b_o$, this can lead to a value
of $m_{3/2}$ of order the weak scale.

\subsection{The View from 11 Dimensions}

It is usually said that the compactification scale,
in string theory, must be comparable to the Planck
mass \cite{vadimscales,couplingsscales}.
The argument is quite simple.
In the heterotic string, the ten dimensional gravitational
and gauge constants are given in terms of the
dimensionless coupling of the theory, which
for now I will denote by $\lambda$, by
\beq
\kappa_{10}^2= \fourth \lambda^2(2 \alpha^{\prime})^4
~~~~~g_{10}^2 = \lambda^2(2 \alpha^{\prime})^3.
\label{heteroticcouplings}
\eeq
Thus the string tension, $T=(2 \alpha^{\prime})^{-1}$
is
\beq
T={g_{10}^2 \over 4 \kappa_{10}^2}={g_{4}^2 \over 4 \kappa_{4}^2}.
\label{tensionis}
\eeq
or
\beq
T \approx 5 \times 10^{17} \rm GeV.
\label{tensionapprox}
\eeq
On the other hand, $g_4^2 = \lambda^2/V T^3$
where $V$ is the volume of the internal space.
So if we require {\it weak string coupling}, $\lambda^2 \le 1$,
$g_4^2 \approx 1$, we have that
$V T^{\prime 3} \approx 1$.  On the
other hand, if $R^{-1} \approx M_{GUT} \approx 2 \times
10^{16}$, this gives
$\lambda^2 \sim 10^7$!

In the past, people have insisted that the coupling
should be less than or of order one, and
have therefore assumed that $R^{-1}$ couldn't be much
different than $M_p$.  However, we have just argued that
string theory should be strongly coupled.  Moreover, in light
of the recent developments in duality, we are not frightened
by the limit $\lambda \rightarrow \infty$:  this should
just be $M$ theory on $X \times S_1/Z_2$, where
$X$ is a six dimensional manifold such as a Calabi-Yau
manifold \cite{horavawitten}. This argument suggests that
$M$ theory might well give a qualitatively better description
of the real world than weakly coupled string
theory \cite{wittenstrong}.

Assuming that the $M$-theory description is valid, one
can determine the $11$-dimensional Planck mass, $M_{11}$
and the size of the $11$'th dimension, $R_{11}$:
\begin{equation}
M_{11}= R^{-1} \left (2 (4 \pi)^{- 2/3} \alpha_{GUT}
\right )^{-1/6}.
\label{meleven}
\end{equation}
\begin{equation}
\Releven^2 =  {\alpha_{GUT}^3 V \over 512 \pi^4 G_N^{2}
 },
\label{rhosquared}
\end{equation}
Substituting $M_{GUT} = R^{-1} = 10^{16}$ GeV, $\alpha_{GUT} = {1\over
25}$ and the correct value for Newton's constant, one finds that
\begin{equation}
R \sim 2 \Meleven^{-1}
\label{melevenr}
\end{equation}
\beq
\Meleven\Releven \sim 72 \ .
\label{melevenreleven}
\eeq
These numbers are quite striking.  $\Meleven$ is not
much different than the unification scale.
The radius of the $11$'th dimension is much larger
than the others, so there is an approximation in
which the universe is five dimensional.
The important scales of new physics are not
set by the usual Planck scale, but in fact
lie at significantly smaller scales.  This observation
has implications for proton decay, since it means
that dangerous dimension five operators are not
suppressed by $M_p$ but rather only by $M_{GUT}$.
Presumably, one would will need approximate
symmetries to account for the smallness of the proton
decay amplitude \cite{bdmtheory}.

In the framework of $M$ theory, we can revisit
the question of the stabilization of the moduli.
In weakly coupled string theories, in addition to the
modulus $S$, one typically speaks of a set of moduli,
which can be thought of as describing the size
and shape of the internal manifold.  To simplify the
discussion, we will speak of one overall size, usually
denoted by $T$ (not to be
confused with the string tension) $T=R^2$, measured in
units of the string tension.
The moduli $S$ and $T$ are now, up to constants:
\beq
S= V_{11} \Meleven^6~~~~~~T= \Releven \Meleven^3 R^2.
\label{mtheoryst}
\eeq
This can be verified by using the relations between
$11$-dimensional and string theory quantities.
Roughly speaking, large $S$ and fixed $T$ now corresponds to large
Calabi-Yau space, while large $T$ and fixed $S$ corresponds to
large radius for the $11$'th dimension.  It is not hard
to work out the Kahler potential for these fields.  One can
do this by first compactifying on the orbifold,
yielding a $10$ dimensional supergravity Yang-Mills theory,
and then reducing on the Calabi-Yau.  This yields
a result identical to that for the ordinary compactification
of string theory at large radius on a Calabi-Yau space.
Now, however, one has an interesting new possibility.
If one calculates the $f$ function at one loop, one finds
that it has the form, for the hidden sector fields \cite{kimchoi,%
ibanezetaloncouplings,wittenstrong},
\beq
f_8 = S- cT\ .
\eeq
This has been calculated both at strong and at weak
coupling; the two computations must -- and do -- agree, by
holomorphy and the periodicity arguments we have
used repeatedly \cite{bdmtheory}. This
result means that the coupling blows up for
$S=cT$ \cite{wittenstrong}. Since in the weakly coupled theory,
$\lambda \rightarrow S /T^3$, this inevitably
implies strong coupling; this is presumably why
this possibility was ignored in the past.  Witten has
suggested that this point, where the coupling
blows up, is special and might
be the location of the true vacuum.  This
assumption yields a prediction
for the ratio of the unification scale and the Planck
scale which is not unreasonable.

While this result is intriguing, it is not easy to go
further.  First, if one examines gluino condensation
in the hidden sector, this tends to drive one away
from the strong coupling point.  Of course, one expects
large corrections as one approaches this point, so the
potential still might have a minimum there.
Second, while this observation
might explain how one linear combination of the
moduli is fixed, it is not easy to see how the other
would be.  Indeed, given that the fifth (eleventh)
dimension is so large, there are important restrictions
on the leading terms in the Kahler potential coming
from five dimensional supersymmetry.  These make
it hard to write any sort of effective theory which
would stabilize the other modulus.   As we did
earlier for the dilaton, we need to suppose that there
are corrections to the Kahler potential for large
$\Releven$ which are surprisingly large \cite{bdmtheory}.
All of this clearly bears further study \cite{horavaoncondensation}.

\section{Conclusions}

Supersymmetry may well be the next layer of structure.
If it is there, and has something to do with the hierarchy problem,
we should see it at the LHC, if not before.  Supersymmetry
is a predictive framework, in that the quantum numbers
and interactions of the array of new particles expected are
well determined.  Much can be said about its phenomenology,
even without a detailed model of the superparticle spectrum.
Strong constraints can be placed on the spectrum by
rare processes.  Unlike the case of, say, technicolor
models, it is not hard to provide models in which all
of these constraints are well satisfied.  Still, one would
like to predict the soft breakings.  This leads to the
crucial question of supersymmetry breaking.

We have explored
two quite distinct ways in which supersymmetry might be
broken.  The first, high energy breaking, raises many puzzles:
why is their squark degeneracy, why is CP-violation so small,
what fixes the numerous parameters of the model?  The second,
low energy breaking, is a much tighter structure, with
few parameters, automatic flavor conservation, and other
desirable features.  Still, it is not at all clear which
structure might ultimately be correct.  No beautiful and
compelling model of low energy breaking yet exists,
and the $\mu$ problem is particularly troubling in this
framework.  Intermediate scale breaking looks like a much
more likely outcome of string theory, though we can hardly
be said to understand string dynamics well enough to
make any definite statements.

Skeptics often ask why one is so interested in supersymmetry,
and ask at one point one will give up on it.  I hope these
lectures have made clear that supersymmetry is of
interest from many points of view:  the hierarchy problem,
dark matter, string theory, and more.  It is hard to imagine
that nature does not take advantage of such
a rich and beautiful structure.  On the other hand,
as compelling as this set of ideas may sometimes seem,
we should all be skeptics.  The experimental support for
supersymmetry is, at best, extremely slender.   As the mass
scales associated with supersymmetry are gradually pushed
higher, one worries that the original argument for
low energy supersymmetry may soon no longer make sense.
Hopefully, time will tell.  In the meantime, there is much
for theorists to do.  In particular, we would like to know
what string theory predicts for the soft breakings.
If we could make progress on this question, we might
someday be in a position similar to that of gauge theories
in the 1970's, exploring and testing the theory through
purely low energy measurements.  Had we never built
the CERN SPS and higher energy machines, we would
still be convinced of the validity of the standard model.
If we could understand the predictions of string theory
for supersymmetry breaking, we could similarly establish
the theory without having to wait for a Planck-scale
machine.  Perhaps, on that highly
optimistic note, it is time to conclude these lectures.


\section*{References}
\begin{thebibliography}{99}

\bibitem{wessbagger}
J. Wess and
J. Bagger, {\it Supersymmetry and Supergravity}, (Princeton University
Press, Princeton, 1983).

\bibitem{wilson}
L. Susskind, Phys. Rev.
{\bf D20} (1979) 2619

\bibitem{couplingunity}
U. Amaldi, W. de Boer and H. Furstenau,
Phys. Lett. {\bf B260} (1991) 447;
C. Giunti, C.W. Kim and U.W. Lee, Mod. Phys.
Lett. {\bf 16} (1991) 1745; J. Ellis, S. Kelley
and D.V. Nanopoulos, Phys. Lett. {\bf B249} (1990) 441;
{\bf B260} (1991) 131; P. Langacker and M.-X. Luo,
Phys. Rev. {\bf D44} (1991) 817.

\bibitem{damien}
D. Pierce, J.A. Bagger, K. Matchev and R.-J. Zhang,
hep-ph/9606211 and references therein.

\bibitem{banksdixon}
T. Banks and L. Dixon, Nucl. Phys. {\bf B299} (1988) 613.

\bibitem{allowedsoft}
M. Grisaru, Nucl. Phys. {\bf B194} (1982) 65.

\bibitem{topquarkloops}
H.E. Haber and R. Hempfling, Phys. Rev. Lett.
{\bf 66} (1991) 1815; J. Ellis, G. Ridolfi and
F. Zwirner, Phys. Lett. {\bf B262} (1991) 477.
 
\bibitem{ibanezross}
L. Ibanez and G.G. Ross,
Phys. Lett. {\bf 110B} (1982) 215.

\bibitem{agcw}
L. Alvarez-Gaume, M. Claudson and M.B. Wise, Nucl.Phys. {\bf B207}
(1982) 96.


\bibitem{dncalc}
J. Polchinski and M.B. Wise, Phys. Lett. {\bf 125B} (1983) 393.

\bibitem{gaillardlee}
M.K. Gaillard and B.W. Lee, Phys. Rev. {\bf D10} (897) 1974;
M.K. Gaillard, B.W. Lee and J.L. Rosner, Rev. Mod. Phys.
{\bf 47} 1975) 277.

\bibitem{georgi}
H. Georgi, {\it Weak Interactions and Modern
Particle Physics}, (Benjamin-Cummings, 1984).

\bibitem{nirreview}
Y. Nir, in Proceedings of the
SLAC Summer Institute, 1992, SLAC-PUB-5847.

\bibitem{nirseiberg}
M. Dine and A. Nelson, {\sl Phys.\ Rev.} {\bf D48} (1993) 1277.

\bibitem{masiero}
F. Gabbiani, E. Gabrielli,
A. Masiero, L. Silvestrini, ``A Complete Analysis of FCNC and CP
Constraints in General SUSY Extensions of the 
Standard Model,'' Nucl. Phys. {\bf B477} (1996) 321,
hep-ph/9604387.


\bibitem{chargedhiggs}
For a recent discussion and references, see J. Hewett and
J.D. Well, ``Searching for Supersymmetry
in Rare B Decays,'' hep-ph/9610323, and references therein.

\bibitem{bg}
R. Barbieri and G.F. Giudice, ``$b\to s \gamma$ Decay and Supersymmetry,''
hep-ph/9303270, Phys. Lett. {\bf B309} (1993) 86.

\bibitem{theoryrb}
The theoretical literature
on $R_b$ is very large.  See,
for example, P.H. Chankowski and S. Pikorski, ``Chargino
Mass and $R_b$ Anomaly, hep-ph/9603310, Nucl. Phys. {\bf B475}
(1996) 3 and references therein.

\bibitem{rparityviol}
S. Dimopoulos and L.J. Hall,
Phys. Lett. {\bf B207} (1987) 210.

\bibitem{weinbergfive}
S. Weinberg, Phys. Rev. {\bf D26} (1982) 287.

\bibitem{dr}
S. Dimopoulos, S. Raby and F. Wilczek, Phys. Lett. {\bf 112B}
(1982) 133.

\bibitem{wittendsb}
E. Witten, {\sl Nucl.\ Phys.} {\bf B188} (1981) 513.

\bibitem{ads}
I. Affleck, M. Dine and N. Seiberg, {\sl Nucl.\ Phys.} {\bf B241}, 
(1984) 493;    {\sl Nucl.\ Phys.} {\bf B256} (1986) 557.

\bibitem{nrtheorems}
M.T. Grisaru, W. Siegel and M. Rocek,
Nucl. Phys. {\bf B159} (1979) 429.

\bibitem{sv}
M.A. Shifman and A.I. Vainshtein, Nucl. Phys. {\bf B277} (1986)
456; Nucl. Phys. {\bf B359} (1991) 571; L. Dixon, V. Kaplunovsky and J.
Louis, Nucl. Phys. {\bf B355} (1991) 649; H.S. Li and K.T. Mahanthappa,
Phys. Lett. {\bf B319} (1993) 152; M. Dine and Y. Shirman,
``Some Explorations in Holomorphy,'' hep-th/9405155, Phys.
Rev. {\bf D50} (1994) 5389.

\bibitem{seibergnr}
N. Seiberg, ``Naturalness Versus Supersymmetric Nonrenormalization
Theorems,'' hep-ph/9309335, Phys. Lett. {\bf B318} (1993) 469;
K. Intriligator, R.G. Leigh and N. Seiberg, ``Exact Superpotentials
in Four Dimensions,'' hep-th/9403198,
Phys. Rev. {\bf D50} (1994) 1092.

\bibitem{dtermnr}
W. Fischler, H.P. Nilles, J. Polchinski, S. Raby and L.
Susskind, Phys. Rev. Lett. {\bf 47} (1981) 757.

\bibitem{newissues}
E. Witten, Nucl. Phys. {\bf B268} (1986) 79.

\bibitem{dsnr}
M. Dine and N. Seiberg, Phys. Rev. Lett. {\bf 57} (1986) 2625.

\bibitem{instanton}
I. Affleck, M. Dine and N. Seiberg, Nucl. Phys. {\bf B241}
(1984) 493; Phys. Rev. Lett. {\bf B51} (1026) 1983.

\bibitem{wittendsba}
E. Witten, Nucl. Phys. {\bf B202} (1982) 253.

\bibitem{dnns}
M. Dine, A. Nelson, Y. Nir
and Y. Shirman,
New Tools for Low Energy Dynamical Supersymmetry Breaking,
hep-ph/9507378,
{\sl Phys. Rev.} {\bf D53} (1996) 2658.

\bibitem{intt}
K. Intriligator and S. Thomas, ``Dynamical Supersymmetry
Breaking on Quantum Moduli Spaces,'' hep-th/9603158,
Nucl. Phys. {\bf B473} (1996) 121.

\bibitem{newmodels}
P. Pouliot,
Phys. Lett. {\bf B367} (1996) 131, hep-th/9510148;
P. Pouliot and M.J. Strassler,
Phys. Lett. {\bf B375} (1996) 175, hep-th/9602031; T. Kawano,
YITP-96-5, hep-th/9602035; E. Poppitz, Y. Shadmi and S.P. Trivedi,
EFI-96-15, hep-th/9605113; EFI-96-24,
hep-th/9606184; K-I. Izawa and T. Yangida, Prog. Theor.
Phys. {\bf 95} (1996) 829, hep-th/9602180;  C. Csaki, L. Randall and W.
Skiba, MIT-CTP-2532; hep-th/9605108; C. Csaki, L. Randall,
W. Skiba and R.G. Leigh, MIT-CTP-2543, hep-th/9607021.

\bibitem{iss}
K. Intriligator, N. Seiberg and S.H. Shenker, ``Proposal for
a Simple Model of Dynamical SUSY Breaking,'' hep-ph/9510203,
Phys. Lett. {\bf B342} (1995) 152.

\bibitem{nelsonnew}
A.E. Nelson, ``A Viable Model of Dynamical Supersymmetry
Breaking in a Hidden Sector,'' hep-ph/9508367, Phys. Lett.
{\bf B369} (1996) 277.

\bibitem{yuri}
Y. Shirman, ``Dynamical Supersymmetry Breaking
Versus Runaway Behavior in Supersymmetry Gauge Theories,''
hep-th/9608147.

\bibitem{cremmer}
E. Cremmer, B. Julia, J. Scherk, S. Ferrar, L.
Girardello and P. von Nieuvenhuizen, Nuclear Physics
{\bf B147} (1979) 105.

\bibitem{nilles}
H.P. Nilles, Phys. Rept. {\bf 110} (1984) 1.

\bibitem{dn}
M. Dine and A.E. Nelson, ``Dynamical Supersymmetry Breaking
at Low Energies,'' hep-ph/9303230, Phys. Rev. {\bf D48} (1993)
1277.

\bibitem{dn2}
M. Dine and A.E. Nelson, ``Low Energy Dynamical Supersymmetry
Breaking Simplified,'' hep-ph/9408384, Phys. Rev. {\bf D51}
(1995) 1362.

\bibitem{bagger}
J. Bagger, ``Weak Scale Supersymmetry:
Theory and Practice,'' hep-ph/9604232.

\bibitem{flavorsymmetries}
This subject has been widely studied.  Some relevant papers are:
M. Leurer, Y. Nir, and N. Seiberg, Nucl. Phys. {\bf B420} (1994)
468; M. Dine, R. Leigh, and N. Kagan, Phys. Rev. {\bf D48} (1993)
4269; D.B. Kaplan and M. Schmaltz, Phys. Rev.
{\bf D49} (1994) 3741; L.J. Hall and M. Murayama, Phys. Rev. 
Lett. {\bf D48} (1993) 4269.

\bibitem{arnowitt}
A.H. Chamseddine, R. Arnowitt and P. Nath, Phys.\ Rev.\ Lett.
{\bf 49} (1982) 970.

\bibitem{agpw}
L. Alvarez-Gaume, J. Polchinski, M.B. Wise, Nucl. Phys. {\bf B221} (1983)
495.

\bibitem{pomarol}
G. Dvali, G.F. Giudice and A. Pomarol, ``The Mu Problem in Theories
with Gauge Mediated Supersymmetry Breaking,'' hep-ph/9603238.

\bibitem{higherorder}
S. Dimopoulos, G.F. Giudice and
A. Pomarol, ``Dark Matter in Theories of Gauge
Mediated Supersymmetry Breaking,'' hep-ph/9607225.

\bibitem{ddrt}
S. Dimopoulos, M. Dine, S.
Raby and  Thomas,``Experimental Signatures of
Low-Energy Gauge Mediated
Supersymmetry Breaking,''
hep-ph/9601367,
Phys.\ Rev.\ Lett. {\bf 76} (1996) 3494.

\bibitem{stw}
D.R. Stump, M. Wiest and C.P. Yuan, ``Detecting a Light
Gravitino at Linear Collider to Probe the SUSY Breaking
Scale, hep-ph/9601362. 

\bibitem{cdfdpf}
D. Toback (for the CDF Collaboration),
``The Diphoton Missing $E_t$ Distribution
at CDF,'' FERMILAB-CONF-96-240-E.

\bibitem{paigeetal}
H. Baer, C.-h Chen, F. Paige and X. Tata,
Phys. Rev. {\bf D49} (1994) 3283.

\bibitem{scottetal}
S. Dimopoulos, S. Thomas and J.D. Wells, ``Sparticle
Spectroscopy and Electroweak Symmetry Breaking
with Gauge Mediated Supersymmetry Breaking,''
hep-ph/9609434; S. Dimopoulos, S. Thomas and
J.D. Wells, ``Implications of Low-Energy Supersymmetry
Breaking at the Tevatron,'' hep-ph/9604452, Phys. Rev.
{\bf D54} (1996) 3283.


\bibitem{dns}
M. Dine, Y. Nir and Y. Shirman, ``Variations on Minimal
Gauge Mediated Supersymmetry Breaking,''
hep-ph/9607397.

\bibitem{dinefischler}
M. Dine and W. Fischler, Phys. Lett. {\bf B110} (1982) 227.

\bibitem{nappiovrut}
C.R. Nappi and B.A. Ovrut, Phys. Lett. {\bf 113B} (175) 1982.

\bibitem{seibergduality}
N. Seiberg, ``Electric-Magnetic Duality
in Supersymmetric Nonabelian Gauge Theories,''
hep-th/941149, Nucl. Phys. {\bf B435} (1995) 129.

\bibitem{ckn}
A.G. Cohen, D.B. Kaplan and
A.E. Nelson, ``The More Minimal Supersymmetric
Standard Model,'' hep-ph/9607394.


\bibitem{cosmostring}
R. Rohm,  Nucl. Phys. {\bf B237} (1984) 553.

\bibitem{narain}
K.S. Narain, Phys. Lett. {\bf 169B} (1986) 41;
K.S. Narain, M.H. Sarmadi and E. Witten,
Nucl. Phys. {\bf B279} (1987) 369.

\bibitem{rohmwitten}
R. Rohm and E. Witten, Ann. Phys. {\bf 170} (1986) 454.

\bibitem{coping}
T. Banks and M. Dine,
``Coping with Strongly Coupled String Theory,''
hep-th/9406132, Phys. Rev. {\bf D50}
(1994) 7454.

\bibitem{strongcoupling}
M. Dine and N. Seiberg,
Phys. Lett. {\bf 162B}, 299 (1985),
and in {\it
Unified String Theories}, M. Green and D. Gross, Eds. (World Scientific,
1986).

\bibitem{shifman}
M. Shifman, ``Little Miracles of
Supersymmetric Evolution of Gauge Couplings,''
hep-ph/96006281.

\bibitem{ibanezlust}
I.E. Ibanez and D. Lust, Nucl. Phys.
{\bf B382} (1992) 305;

\bibitem{vadim}
V. Kaplunovsky and J. Louis, Phys. Lett. B
{\bf 306} (1993) 269.


\bibitem{wittendr}
E. Witten, Phys. Lett. {\bf 155B} (1985) 151.

\bibitem{racetrack}
V. Krasnikov, Phys. Lett. {\bf 193B} (1987)
37; L. Dixon, in the {\it Proceedings of the
DPF Meeting, Houston, 1990}; J.A. Casas,
Z. Lalak, C. Munoz and G.G. Ross,
Nucl. Phys. {\bf B347} (1990) 243;
T. Taylor, Phys. Lett. {\bf B252} (1990) 59.


\bibitem{louisnir}
J. Louis and Y. Nir, ``Some Phenomenological Implications of
String Loop Effects,'' hep-ph/9411429, Nucl. Phys. {\bf B447} (1995) 18.

\bibitem{louisdilaton}
R. Barbieri, J. Louis and
M. Moretti, ``Phenomenological Implications of Supersymmetry
Breaking by the Dilaton,'' hep-ph/9305262, Phys. Lett. {\bf B312}
(1993) 451, Erratum -- ibid {\bf B316} (1993) 632.

\bibitem{vadimscales}
V. Kaplunovsky, Mass Scales of String
Unification, {\sl Phys. Rev. Lett.} {\bf 55} (1985) 1036.

\bibitem{couplingsscales}
M. Dine and N. Seiberg,
Couplings and 
Scales in Superstring Models, {\sl Phys. Rev. Lett.} {\bf 55}
(1985) 366.

\bibitem{horavawitten}
P. Horava and E. Witten, ``Heterotic and Type I String Dynamics From
Eleven Dimensions'', {\it Nucl. Phys.} {\bf B460}, (1996) 506, 
hep-th/9510209.

\bibitem{wittenstrong}
E. Witten, ``Strong Coupling Expansion of Calabi-Yau Compactification,''
hep-th/9602070, Nucl. Phys. {\bf B471} (1996) 135.

\bibitem{bdmtheory}
T. Banks and M. Dine, ``Couplings and Scales in
Strongly Coupled Heterotic String Theory", hep-th/9605136.

\bibitem{kimchoi}
K. Choi and J.E. Kim, {\it Phys. Lett.} {\bf 165B}, (1985), 71.

\bibitem{ibanezetaloncouplings}
L. Ibanez and
H.P. Nilles,  Phys. Lett. {\bf 180B} (1986), 354.

\bibitem{horavaoncondensation}
P. Horava, ``Gluino Condensation in Strongly Coupled
Heterotic String Theory,'' hep-th/9608019.


 \end{thebibliography}
\end{document}